\documentclass[acmsmall]{acmart}

\AtBeginDocument{%
 }

\setcopyright{rightsretained}
\acmJournal{PACMHCI}
\acmYear{2024} \acmVolume{8} \acmNumber{CSCW2} \acmArticle{} \acmMonth{11} \acmPrice{}\acmDOI{XX.XXXX/XXXXXXX}

\usepackage{xcolor}
\usepackage{soul}
\usepackage{comment}
\usepackage{lipsum}

\begin{document}

\title[Studying Up Public Sector AI]{Studying Up Public Sector AI: How Networks of Power Relations Shape Agency Decisions Around AI Design \& Use}


\author{Anna Kawakami}
\affiliation{%
  \institution{Carnegie Mellon University}
  \city{Pittsburgh}
  \state{PA}
  \country{USA}
}
\author{Amanda Coston}
\affiliation{%
  \institution{Carnegie Mellon University}
  \city{Pittsburgh}
  \state{PA}
  \country{USA}
}

\author{Hoda Heidari}
\authornote{Co-senior authors contributed equally to this research.}
\affiliation{%
  \institution{Carnegie Mellon University}
  \city{Pittsburgh}
  \state{PA}
  \country{USA}
}

\author{Kenneth Holstein}
\authornotemark[1]
\affiliation{%
  \institution{Carnegie Mellon University}
  \city{Pittsburgh}
  \state{PA}
  \country{USA}
}

\author{Haiyi Zhu}
\authornotemark[1]
\affiliation{%
  \institution{Carnegie Mellon University}
  \city{Pittsburgh}
  \state{PA}
  \country{USA}
}

\renewcommand{\shortauthors}{Anna Kawakami et. al.}

\begin{abstract}
As public sector agencies rapidly introduce new AI tools in high-stakes domains like social services, it becomes critical to understand \textit{how} decisions to adopt these tools are made in practice. We borrow from the anthropological practice to ``study up'' those in positions of power, and reorient our study of public sector AI around those who have the power and responsibility to make decisions about the role that AI tools will play in their agency. Through semi-structured interviews and design activities with 16 agency decision-makers, we examine how decisions about AI design and adoption are influenced by their \textit{interactions with} and \textit{assumptions about} other actors \textit{within} these agencies (e.g., frontline workers and agency leaders), as well as those \textit{above} (legal systems and contracted companies), and \textit{below} (impacted communities). By centering these networks of power relations, our findings shed light on how infrastructural, legal, and social factors create barriers and disincentives to the involvement of a broader range of stakeholders in decisions about AI design and adoption. Agency decision-makers desired more practical support for stakeholder involvement around public sector AI to help overcome the knowledge and power differentials they perceived between them and other stakeholders (e.g., frontline workers and impacted community members). Building on these findings, we discuss implications for future research and policy around actualizing participatory AI approaches in public sector contexts.
\end{abstract}




\maketitle
%\hhcomment{I wonder if it makes sense to replace ``design and use'' in the title with ``adoption''. I don't think we say anything specifically about ``design'' hence the suggestion. Also as additional keyword, we can add ``AI procurement'' or ``AI adoption''.}\akcomment{I'll add the procurement term in key works - thanks for the suggestion! And I think we say things about design in the findings (e.g., workers wanted to inform design at earlier stages.}
\section{Introduction}
% 1 - Public sector is rapidly deploying AI tools but signs of success have been more unclear (a range of real-world challenges) 

 % \khcomment{\textbf{General note for proofreading passes (feel free to expand):} \\ make sure topic sentences and headers are clear, and that they capture what follows in a given paragraph or section.}

Public sector agencies across the United States are rapidly deploying AI-based tools to assist or automate complex, high-stakes decisions traditionally made by frontline workers in areas such as social services, public health, criminal justice, and education~\cite{barocas-hardt-narayanan, Levy2021,yu2018artificial,veale2018fairness}. Agencies have been motivated to explore the use of these new AI tools to help overcome resource constraints and limitations in human decision-making (e.g., biases in worker decisions)~\cite{tan2018investigating,bansal2019beyond,chouldechova2018case}. However, the use of AI tools in these domains have been met with significant contention. A growing body of research and public media discusses issues of bias~\cite{obermeyer2019dissecting,chouldechova2018case}, lack of transparency ~\cite{saxena2020human,Brown2019,zytek2021sibyl}, and lack of validity~\cite{coston2022validity} of public sector AI deployments, amongst other concerns. %In turn, in the past years, public sector agencies have often dropped AI tools after they failed to produce the value it was intended to produce in practice~\cite{}. 
To address these concerns, the human-computer interaction (HCI) and machine learning (ML) communities are increasingly exploring approaches to improve the design and use of public sector AI tools. For example, prior work on AI-based decision support tools has studied racial disparities in algorithmic predictions and ways to mitigate them ~\cite{obermeyer2019dissecting,cheng2022child,holstein2019fairness,VandenBroek2020}, interface-level interventions for frontline workers using these tools~\cite{kawakami2022care,holstein2017intelligent,HoltenMoller2020,kang2022stories,Yang2019}, and community members' concerns and desires around them~\cite{Brown2019}. This work has substantially advanced our understanding of the social impacts of public sector AI systems, and opportunities for improving how they are designed and used.
% note to self -- can tighten the above, based on what you say here. 

The challenges outlined by prior work point to a crucial, yet under-examined question: How do public sector agencies make decisions about \textit{whether} to create or adopt a new AI tool in the first place? While prior theoretical work has established the consequential nature of decisions and assumptions made in early stages of model design (e.g., in problem formulation~\cite{wang2022against, passi2019problem}), we still lack an empirical understanding of \textit{how} these decisions are made in practice. 
% 3 - Most of this work has decentered systems of power  
Addressing this question requires us to turn our attention to actors with significant power to shape AI design, adoption, and use. %\khcomment{Question: Does this framing capture frontline leadership, or would some wording adjustment help to better capture this group? Is frontline leadership included under ``technology evaluators''?} \akdelete{Recent research has urged academic researchers to consider larger \hzedit{social systems where the AI system is embedded} when studying or designing AI~\cite{barabas2020studying,miceli2022studying}. } 
In doing so, we borrow from anthropologist Laura Nader's 1972 call to ``\textit{study up}'' -- reorienting anthropological fields of study around those who have power and responsibility to shape social systems and norms~\cite{nader1972up}. %\akcomment{could include one mor sentence here, maybe quoting Nader}\hzcomment{+1. We probably want to add 1-2 sentences here to highlight the benefits of using the "study-up" lens according to the prior literature.} 
In the study of AI systems, recent work has urged academic researchers to reorient the studies of algorithmic fairness~\cite{barabas2020studying} and machine learning datasets \cite{miceli2022studying} around issues of power, demonstrating the value of studying up technology as a means to uncover broader social and infrastructural challenges impeding the responsible development and use of AI systems. More broadly, literature in the Science and Technology Studies have contributed a range of concepts and frameworks to understand relationships between power and technology (e.g., in~\cite{kitchin2014towards,crawford2021atlas}). We extend these prior lines of work, by grounding an empirical understanding of how power structures impact on-the-ground decisions around public sector AI. %\hhcomment{Add citation to back up this claim}. 
%\hhdelete{Such reorientations have surfaced \hl{...} that have been overlooked in existing studies of AI systems. Extending these calls, we argue that, in the study of public sector AI, overlooking \hl{...} how the design and use of AI are shaped by, or shape, broader \hl{...} \hzcomment{Maybe we no longer need these (incomplete) sentences.}} % Existing design proposals have also inadequately considered how these approaches might be shaped by, or shape, broader ecosystems of power and infrastructure. 

% 4 - We focus on social services. A bit more detail on why this is a compelling domain to study. 
% To fully address how broader systems of power shape the design and use of public sector AI, it is critical to shift our gaze upwards to “study up” the perspectives of those in power~\cite{}. 

In this paper, we 
\textit{study up} public sector AI via the perspectives and experiences of agency decision-makers: those typically occupying upper levels of a public sector agency's organizational hierarchy, holding the power and responsibility to progress, halt, or otherwise shape whether and how AI tools are designed and used. We focus on public sector AI in social services, where ``social services'' refers to a broad range of government services intended to benefit the community, such as housing, education, child welfare, and healthcare.
%\khcomment{Note: I'm adding a brief definition of social services here, which might be especially useful for some non-US-based reviewers (please feel free to update and add a citation)} Social services \khedit{have}\khdelete{is a domain that has} seen rapid advancements in the use of new AI tools, and along with it, a growing trend of AI tools getting dropped after failures are observed post-deployment~\cite{samant2021family}. 
In HCI and CSCW research, much of our knowledge around public sector AI comes from the perspectives and experiences of those directly impacted by higher-up decisions (e.g.,  frontline workers~\cite{saxena2021framework,kawakami2022improving} and community members ~\cite{Brown2019,stapleton2022imagining}). 
%\akdelete{In this study, we reorient our gaze upwards to (re)ask how \hhedit{those in relatively higher positions of power} (e.g., directors of agencies and departments, researchers and developers) in public sector agencies make decisions around the design and deployment of AI tools in social services.} 
To better understand the role of agency decision-makers, we ask the following research questions:
\begin{itemize}
    \item What factors are agency decision-makers' shape decisions to move forward with a proposed AI project? 
    \item What opportunities exist to improve such decision-making?
\end{itemize}
To address these questions, we conduct semi-structured interviews and design activities with 16 agency decision-makers across three public sector agencies in the United States. Our analysis surfaced how \textit{power relations with} and \textit{assumptions about}
% Anna note on edit above - I think Ken had previously mentioned this current framing may lack novelty if it's as part of our results (so I reverted the edit!)
%\khcomment{Clarity: It's not so clear to me, as currently written, what the ``however'' statement is here, and why prior work mapping decision points in healthcare is brought up above. I wonder whether it would be clearer to delete the text from ``Prior work maps ...'' up to ``We therefore organize ...'', and instead just say something like ``Based on our analyses, we organize our findings to center the relationships ...''}
other actors---situated within, above, and below the agency---shaped their decisions around AI design, adoption, and use. In particular, we contribute insights into:
%\khcomment{Very minor question: do we think readers will understand what ``layered tensions and barriers'' means? (particularly the meaning of ``layered'')} \akcomment{hmm... good point. I removed it and highlighted the phrase, to look for alternate phrases later.}

%\khcomment{I wonder whether it could help to include a topic sentence at the start of each of the first two bullet points, which highlights how these findings advance our knowledge (e.g., ``Whereas prior HCI literature has ... we find that ...'' or ``Prior HCI research has tended to focus on ... In this study, we find that ...). (the third bullet already has this!) The rest of the text that follows under each bullet point would then function as } \akcomment{good point! taking a pass at this now.}
% \khcomment{to-do: update headers of these paragraphs for clarity, and take a pass to better highlight novelty}
\begin{itemize}
    \item \textbf{Dynamics and sources of disagreement around AI \textit{within} public sector agencies:} When discussing public sector agencies' decision-making around AI adoption and use, prior academic and grey literature has often treated agencies as monolithic entities. Our findings surface how existing depictions of agencies overlook the intra-agency contention that exists amongst workers around these decisions. Participants described how agency workers--including those with relatively higher institutional power (e.g., agency leaders, scientists) and frontline workers--held dissenting perspectives on the validity and value of AI tools. While these workers with higher institutional power advocated for increased community engagement in their agency and more reflexive deliberation on technical design decisions, their perspectives were sometimes ignored or questioned by their colleagues, the majority of whom were perceived to be interested in advancing their agency's use of AI tools. Agency leaders and developers promoting the value of new AI tools described efforts to mitigate frontline workers' concerns towards AI, which they believed to be especially pervasive and persistent, by ``convincing'' frontline workers of the value of AI tools. Talking with frontline workers who were involved in piloting and providing feedback on new AI tools, we observed disconnects between the reasons underlying frontline workers' concerns and the reasons often attributed to them by agency leaders and developers. (Section~\ref{within})
    \item \textbf{How pressures from external institutions shape decisions around AI design, adoption, and use:} Our findings suggest that agency leaders and creators of public sector AI tools experience constant pressures from other powerful institutions (e.g., legal systems and state courts, private companies), that shape decisions around AI design, adoption, and use. Whereas prior empirical research has tended to examine the organization as a unit of analysis (e.g., to study organizational governance, worker interaction, or model design), our findings help contextualize how agencies' interactions with external institutions of high power may impact those organizational practices and decisions. For example, participants felt frustrated with their inability to interrogate ethical considerations when adopting AI tools from private companies. They felt constrained by unfavorable working relationships predefined by procurement contracts, which disincentivized private entities from being transparent in their model development and evaluation process. Other participants shared concerns that their AI tools, while intended to be used by frontline workers, could be misused as evidence for or against the agency in legal court cases. Participants described that these legal pressures were further exacerbated by constrained communication channels, including the lack of interactions with legal experts when creating or adopting AI tools. (Section~\ref{above})
     %\khcomment{Clarity: I wonder whether readers will understand what ``experienced ad-hoc encounters'' means here. Is it important to the story that they are ad-hoc (if so, why)? And are they all best described as ``encounters''? Or are some of these better thought of as constant pressures (i.e., as organizations, they are situated in these broader systems of power)?} \akcomment{I tried an edit!}
   %  As a result, participants desired more formalized mechanisms for interacting with them to inform design and deployment decisions and support from other institutions (federal government) to help overcome incentive misalignments with private companies. \khcomment{Clarity: I wonder whether readers will understand what ``formalized mechanisms for interacting with them'' means. Maybe this can be clarified through a small edit/elaboration} (Section~\ref{above})
    \item \textbf{Barriers and disincentives to hearing and empowering impacted communities:} While academic research has proposed a multitude of approaches (e.g.,~\cite{pilemalm2018participatory,Smith2020}) to expand stakeholder involvement in AI design and development, most participants perceived infrastructural and communication barriers between them and impacted community members external to their agency; these barriers created additional disincentives for connecting with or involving impacted communities in their AI design process.  %\khcomment{Novelty: It seems like the lack of (meaningful) stakeholder involvement throughout the design process of public sector AI has been fairly well documented in prior literature. In other words, the observation of this gap between research and practice may not be surprising to CSCW readers on its own. I wonder whether there's a way to better highlight new insights in the preceding sentences.} 
    Participants--who were experts in leading agencies or developing technology but had minimal to no experience in community outreach--felt they were ill-equipped to make decisions about how to involve the community: which groups to reach out to, how to develop sustainable relationships with them, and how to mitigate power imbalances when involving them. As a result, participants often described that interactions with the communities they serve were either very limited, heavily constrained, or non-existent. (Section~\ref{below})
    %\akcomment{Add this to the above: " Finally, we expand on the infrastructural and communication barriers participants perceive between their agency and the communities they serve, that further disincentivize agencies from expanding community participation in their AI design processes (Section~\ref{below})." Reminder to also add this to Sec 4.3 introduction.}
\end{itemize}
%\akdelete{Overall, our findings surface how a broader ecology of systems (legal, social, infrastructural) and stakeholders (both internal and external to agencies) \khdelete{that }collectively shape public sector agencies' creation and use of AI systems in social services.} 
% \akdelete{\hhdelete{Overall, by}\hhedit{By} centering the network of power relations public sector agencies operate within, our findings contribute to and help contextualize the growing literature describing real-world socio-organizational factors impacting the design and use of AI systems in high-stakes, public sector domains (e.g.,~\cite{kawakami2022improving,saxena2021framework,kim2023organizational}).}
By studying up the network of power relations public sector agencies operate within, we advance an empirical understanding of current challenges to actualizing AI development improvements in practice, even when argued as best practices in the research literature. While agency decision-makers hold direct responsibility for making the decisions that legitimize existing practices, our findings surface a range of legal, infrastructural, and social barriers to responsible design that agencies are currently ill-equipped to address alone. Developing the systemic conditions and practical resources needed to shift towards an improved practice requires urgent attention from the research and policy communities. In Section~\ref{sec:discussion}, we discuss opportunities to make this shift through interventions that streamline the adoption and implementation of participatory AI approaches, specifically in public sector contexts. Towards this goal, we discuss how research and policy interventions can reshape--as well as leverage--the network of power relations that public sector agencies reside within.

\section{Background and Related Work}
\subsection{AI in Public Sector Social Services}
% Which domains AI has been used in social services, and findings around the use and design of AI in social services, using child welfare as an example 
As public sector agencies aim to mitigate resource constraints and improve the decision-making quality of their staff, several have turned to exploring the use of new AI tools. In social services alone, AI tools have been rapidly deployed to assist decisions around child maltreatment screening~\cite{chouldechova2018case}, homeless services~\cite{kuo2023understanding}, education~\cite{holstein2018student,robertson2021modeling}, and predictive policing~\cite{shapiro2017reform}. However, the deployment of AI tools in social services, and especially predictive AI tools, has also introduced a myriad of ethical and social concerns. For example, researchers have raised questions around systemic biases embedded in AI tools~\cite{cheng2022child, eubanks2018automating}, the validity of the models used~\cite{coston2022validity,raji2022fallacy}, and the (lack of) support for frontline workers using them~\cite{kawakami2022improving}. 

Towards mitigating the potential social harms of AI tools introduced to the public sector, prior work has proposed various approaches to improve technical design decisions. Acknowledging the value-based tensions across the many stakeholders of AI tools~\cite{HoltenMoller2020}, many of these approaches aim to incorporate stakeholder-specific values and knowledge into the design of algorithms, for example, through expanding participation along the AI design process. Prior work has proposed a myriad of participatory approaches~\cite{delgado2021stakeholder} (e.g., value-sensitive algorithm design ~\cite{zhu2018value}, deliberation-based participatory algorithm design ~\cite{zhang2023deliberating}) to solicit and operationalize stakeholders’ desires and values into the design of algorithmic systems. Such work has often also surfaced the challenges of actually incorporating multiple, conflicting values into an algorithm. For example, Møller et. al.~\cite{HoltenMoller2020} discusses deviations in what notions of value that developers and caseworkers believed should be considered in metrics for algorithmic decision-making systems.  

% Towards improving the integration of technology in social services, a growing body of literature has aimed to shed light onto how AI tools are integrated into agencies. 
% Much existing work around how AI tools are \textit{actually} introduced into agencies, and in turn, how improved design approaches may get integrated into real-world context
% has focused on understanding the experiences and needs of AI developers who build and frontline workers who use these tools (e.g., ~\cite{saxena2021framework,kawakami2022improving}). 
A growing body of literature has shed light on how AI tools are actually integrated into social service agencies (e.g., ~\cite{saxena2021framework,kawakami2022improving}),  implicating how design approaches and interventions proposed in the academic literature may, in the future, \textit{actually} fit into real-world contexts~\cite{kawakami2024situate}. This work has begun to surface additional challenges arising from sources beyond the model--for example, from organizational pressures and incentives, or power imbalances across stakeholders~\cite{kawakami2022improving}. These findings raise questions around how organizational and social factors may shape practices and norms surrounding the design and deployment of public sector AI. This work has largely relied on the insights of AI developers who build and frontline workers who use AI tools, leaving open the question how those holding positions of higher power and responsibility, like directors and managers in public sector agencies, influence these decisions. Our work aims to illuminate the perspectives and experiences of those holding positions of higher power who ultimately drive decisions around the design and deployment of AI. 

\subsection{Studying Up Algorithmic Systems}
% One paragraph overviewing eixsting work on studying up technology (can change the above to AI, to narrow the scope) 
Towards better understanding the broader systems of power and privilege shaping the impacts of algorithmic systems, a growing chorus of researchers have called for a reorientation of the study of AI around issues of \textit{power}~\cite{barabas2020studying,miceli2022studying,seaver2014studying,forsythe2001studying}. Prior work has examined the downstream consequences of decisions made by those who develop and govern AI technologies, to examine tensions between impacted stakeholders and institutions of power (e.g.,~\cite{dow2018between}). 

Drawing on anthropological practices to study those in higher positions of power, researchers have increasingly recognized that looking downwards only provides a partial understanding of the broader social and structural mechanisms that shape AI systems. Recent work has called for more researchers to shift their gaze upwards, to study the practices of more powerful stakeholders. These calls have reflected on the relevance of Laura Nader's 1972 proposal to  ``\textit{study up}''~\cite{nader1972up}, urging anthropologists to reorient their studies to examine how the practices of powerful institutions and authority shape everyday experiences and norms. In recent years, Barabas et. al.~\cite{barabas2020studying} discussed the need for a similar reorientation around power in the study of data science, explaining: ``The political and social impacts of algorithmic systems cannot be fully understood unless they are conceptualized within larger institutional contexts and systems of oppression and control.'' Miceli et. al.~\cite{miceli2022studying} further extend these calls to re-examine machine learning data issues typically labeled as problems of ``bias,'' demonstrating how viewing them as problems of power surfaces often overlooked factors around social contexts (e.g., labor conditions, epistemological stances) underlying data problems.   

To holistically understand the social mechanisms shaping \textit{how} AI systems could bring downstream harm to vulnerable communities, we extend this line of work, revisiting questions around the design and deployment of AI systems in the public sector. We shift our gaze upwards to focus on the decisions and assumptions made by stakeholders with relatively high amounts of decision-making power in public sector agencies, and the broader social systems and infrastructures surrounding them that shape their decisions around AI.

\section{Methods}
To ``study up'' the perspectives of those in positions of power to shape the design and deployment of public sector AI, we conducted semi-structured interviews and design activities with 16 participants across three public sector agencies. %\khcomment{Note on the previous sentence (may also be relevant for the intro): I wonder whether "broader systems that shape the public sector AI design and deployment processes" is the best way to describe what we study in this work. Perhaps useful to scope down this statement around the group of participants included in the study?} \akcomment{thanks for your note! I tried an edit above.}
We recruited agency stakeholders who have experience making or shaping decisions around the design and deployment of public sector AI technologies. In the following, we describe our recruitment process, study procedure, and analysis method. 
% six in leadership or managerial positions, six in developer positions, four in the frontline (including those in managerial positions and those working on the ground)
\subsection{Public Sector Agencies and Participants} \label{participants}
To understand perceptions, practices, and challenges shaping the design and deployment of public sector AI across contexts, we aimed to recruit participants from a range of public sector agencies across the United States. Our team searched for U.S.-based state, city, or county-level public sector agencies with human service departments (e.g., child welfare, predictive policing) that have previously or are currently considering the use of AI tools, had previously deployed then stopped using AI tools, and/or are currently using AI tools. We found 19 agencies fitting these criteria based on information available on public agency websites or news articles. At each of these agencies, we found contacts in leadership positions (e.g., directors of departments), and emailed them stating our research goals and interest in starting a conversation. Out of the 19 agencies, contacts from four of the agencies responded and agreed to having an initial conversation. Out of those four agencies, three agencies agreed to participate in our study. 

From these three agencies, we aimed to recruit participants in occupations that are typically tasked with making or informing decisions around the use or design of AI tools. On aggregate, we interviewed directors of human service departments, directors and managers of teams responsible for developing AI tools, and researchers and analysts who are involved in building or analyzing AI tools. In one agency, our contact connected us to frontline staff who played a role in piloting AI tools or had otherwise also discussed the use of AI tools with developers and leaders in their agency. In our communication to each agency, we requested study participation from the same set of occupations. The actual occupations and numbers of participants involved in our study were determined by who our contacts at the agencies connected us to. 

We refer to participant occupations using four stakeholder groups: \textbf{1}) \textit{Agency Leaders} (L): Individuals in director or managerial roles who typically are involved in making agency- or department-level decisions, \textbf{2}) \textit{Research and Development Workers} (RD): Individuals in research, development, and/or analysis teams internal to a given public sector agency who are involved in the creation or evaluation of AI tools used by the agency, \textbf{3}) \textit{Frontline Leaders} (FL): Individuals who work in leadership roles within human service departments in a public sector agency and who typically are in closer communication with frontline workers than Agency leaders, and \textbf{4}) \textit{Frontline Workers} (FW): Individuals at a public sector agency whose occupations bring them in direct contact with the community their agency serves. Our study includes six Agency Leaders, six Research and Development Workers, two Frontline Leaders, and two Frontline Workers. %Table~\akcomment{table ref} includes aggregated participant demographics. 

\subsection{Semi-Structured Interviews and Design Activity}
We conducted 90 minute study sessions with each session including a semi-structured interview and design activity. The semi-structured interview included three sections. First, we asked participants about their background, for example, questions related to their current role in the public sector agency. Next, we asked participants about their current and past practices for making decisions around whether to design, deploy, or adopt a given AI tool in their public sector agency. This involved asking questions about what considerations aided these decisions (e.g., goals for introducing algorithm, empirical evaluations conducted). In the next section, we probed more deeply on challenges they faced in past experiences designing, deploying, or adopting AI tools. 

To complement our semi-structured interview, we also conducted a design activity. Through the design activity, we invited participants to ideate future considerations that they believed should aid decisions around the design, deployment, or adoption of public sector AI tools. We asked follow-up questions to better understand participants' past experiences and challenges that shaped their ideas. By prompting participants to ideate considerations for a hypothetical future state, we opened opportunities for participants to also reflect on and identify a broader set of challenges they currently face that otherwise may not have surfaced during the semi-structured interview. For example, participants who described a future need to expand community participation in the AI design process also reflected on current barriers their agency faces to supporting external participation in their design process. In this case, because community participation does not currently exist at the participants' agency, participants seldom brought up these barriers during the semi-structured interview, which was more focused on understanding current practices and challenges. To support the design activity, we shared our screen and shared a link to an online board on Figma, a collaborative web application that accommodates multiple users. As participants ideated future considerations they believed should aid the hypothetical design and deployment of AI technologies, we used online post-its to help document and organize their ideas. 

All participants completed the design activity. Participants who were not involved in decisions around whether to design and deploy AI tools (i.e., frontline managers and frontline workers) moved onto the design activity after completing the first portion of the semi-structured interview (on participant background). Participants who participated in all three sections of the semi-structured interview completed the design activity immediately afterwards. All 16 participants consented to having their audio recorded, shared screen recorded, and having notes from the design activity recorded and saved. We requested individual study participation from participants. Some participants preferred to participate in the research study in pairs because they felt that the pair would provide a more extensive set of insights together, or because they wanted to provide the other participant an opportunity to experience an externally conducted research study. Thus, we conducted the semi-structured and design interviews either individually or with pairs of participants.

\subsection{Analysis}
The ~16 hours of interview recordings were transcribed then qualitatively coded using reflexive thematic analysis~\cite{braun2019reflecting} by two researchers. We ensured that all interviews were coded by the first author who conducted all the interviews, and, whenever applicable, another author who observed the interview. The first author coded one transcript first, then discussed the codes with others to align on coding granularity. Each coder prioritized coding observations related to the research questions while also remaining open to capturing a broader range of potential findings. We resolved disagreements between coders through discussion. 

After coding the transcripts, we conducted a bottom-up affinity diagramming process~\cite{beyer1999contextual}. Through multiple rounds of refining and grouping, we iteratively grouped 506 unique codes into three levels of themes. The first level grouped the 506 codes into 52 themes; the next level grouped these 52 themes into 17 second-level themes; and the last level grouped these 17 themes into the three highest level themes. We organized the Findings (Section~\ref{findings}) by these three highest level themes. 

\subsection{Ethics and Participant Safety}
We assured all participants that their participation is voluntary, all study questions were optional, their responses will remain anonymous, and they may ask us to remove content at any time after the study. To mitigate the risk that participants are identified within their agency and that agencies are identified, we do not specify participant roles beyond the participant groupings we define in Section~\ref{participants}. To mitigate the risk that public sector agencies are identified, we do not differentiate the agencies that participants are employed at and limit the amount of details we provide about each agency. This study was approved by the Institutional Review Board of the authors' home institutions.

\subsection{Positionality}
Our research team collectively holds expertise across human-computer interaction, computer-supported cooperative work, critical computing, public policy, machine learning, and statistics. As researchers studying public sector settings, we are conscious of our position of privilege in having research access to public sector agencies and leaders, who have opened their doors to us as external researchers. We acknowledge that our positionality and direct interactions with these agencies, along with our prior work examining the downstream impacts of public sector AI systems on social workers~\cite{kawakami2022care,kawakami2022improving,cheng2022child} and community members~\cite{stapleton2022imagining}, have collectively shaped our perspectives which guide our approach to this research.
% Reminder: cite our own papers on community and frontline perspectives 

\section{Findings} \label{findings}
%\khcomment{To strengthen the introduction to the findings section, it would help to update this first paragraph to more explicitly, briefly summarize \textit{how} these findings progress discourse (and which discourses, among whom). The first sentence of the paragraph could then be removed. For example, this might involve statements like ``Whereas prior HCI research on public sector AI has ......., we shed light on how .....'' and ``Prior treatments of ...... in research and popular media suggest that ........ However, our findings ............''. Note: Some of this is already here, but may get a bit buried in the second paragraph.} 
% Whereas prior HCI research on public sector AI has ......., we shed light on how ..... and 
As public sector agencies continue to develop and drop AI tools deployed in high-stakes decision-making domains, research and popular media has increasingly scrutinized their decisions~\cite{samant2021family,wang2022against,gerchick2023devil}. Agencies have typically been depicted as monolithic entities making authoritative and self-assured decisions around AI. However, our findings complicate this depiction. 

We unveil how agency decisions around AI design and use are shaped by agency-internal contention regarding the value and validity of AI, constant pressures from other institutions of power, and concerns around their ability to mitigate power and knowledge differentials between them and impacted communities. To contextualize agency decisions within the broader web of infrastructural and social systems they reside in, we organize our findings to center the power relations that public sector agencies hold with other sectors and stakeholders situated \textit{within} (Section~\ref{within}), \textit{above} (Section~\ref{above}), and \textit{below} (Section~\ref{below}) the agency.
% We explore the impact of agencies' relationships with various stakeholder communities on their decision-making processes. In particular, we describe how these relationships--including with external legal systems, contracted private companies, and impacted communities, and internal agency workers--play a critical role in shaping whose concerns are heard, whose expertise is valued, and which practices become normalized as public sector agencies continue to develop and use new AI tools. 
%For each subsection, we discuss implications and opportunities for improving public sector decisions around the design and deployment of AI. %We show a summary of findings and implications in Table n. %\hl{Overall, our findings suggest that, without adequate consideration of these ..., future ... missed opportunities}.
% Guidelines are sometimes unrealistic, and more incentives are needed for adoption (e.g., regulation)
\subsection{Power Relations within the Agency: Misconstrued Concerns and Tensions around AI amongst Workers} \label{within}
Agency decision-makers' experiences surfaced contrasting viewpoints around the value and validity of AI among different groups of workers, including within frontline workers, within Research and Development (R\&D) workers, and even within agency directors. Agency decision-makers who were proponents of increasing their agency's use of AI tools (who were the majority of the participants) felt concerned about their peers' potential misunderstandings on the role that AI would play in their agency, especially when these concerns led to the suspension of certain AI tools (Section~\ref{within_value}). They were especially wary of frontline workers' seemingly pervasive and persistent concerns towards AI. As a result, agency decision-makers described a range of efforts to promote an organizational culture of understanding of AI, believing that disclosing how its agency would and would not use AI could help mitigate concerns. However, talking with frontline workers who were involved in sharing feedback on piloted AI tools, we observed critical gaps between frontline workers' underlying concerns and the reasons for concern that agency leaders and developers attributed (Section~\ref{frontline_concerns}). Moreover, while agency decision-makers described that they involved frontline workers in decisions around AI design and use, frontline workers expressed that their existing pathways to inform decisions were too constrained; they desired opportunities to inform key decisions at earlier design stages (Section~\ref{frontline_desires}).

\subsubsection{\textbf{The value of AI tools is a point of contention among workers in agencies.}}\label{within_value}
% How workers within agencies disagree about whether and when to use AI.
% \accomment{I changed the focus from disagreement to concerns because right now the quotes all highlights concerns raised within the agency. If we want to focus on disagreement, then we should have some discussion and quotes showing support of algorithms.  }
While public sector agencies are often portrayed in the media as collective forces pushing forth the deployment of new AI tools~\cite{samant2021family}, our findings suggest this depiction overlooks critical internal tensions and disagreements between agency stakeholders -- including those with the power and responsibility to shape the design of these tools.

Recalling past experiences where an individual's opposition towards AI tools led to deployment halts, several Agency Leaders and R\&D Workers described that one of their biggest challenges to moving forward with the use of AI-based tools was receiving ``buy-in'' from other leaders in their agency. This was particularly important to progressing innovations on AI in the agency, because, historically, different leaders across time and within the same agency held divergent stances around the value and validity of AI tools. Because of the positions of power that these leaders held, their stances towards AI had dramatically impacted decisions around whether to deploy AI tools within their agency. 
% \khcomment{Is it interesting/notable that buy-in is required from leadership? If so, it could help to add a sentence or so more to explain what may be interesting about this observation. Or is this a minor point that you want to cover really quickly, so that you can get to more interesting findings below? ...Hm, actually, I wonder whether an interesting point to highlight (which is not currently coming through here) is that different leaders (across different agencies, or across time in the same agency) can have very different stances, and that these differences can dramatically impact decision-making around AI deployments?} 
For example, during the interview, one participant recalled an experience where their agency had developed a new AI tool that outputs risk scores, intended to assist frontline workers' screening decisions around potential child maltreatment cases. The participant, who managed a team of AI developers in their agency, described that this tool was never deployed due to opposition from an agency director: 
\begin{quote}
% ``there's no particular reason why [one AI tool is] actually deployed and operationalized, and [the other AI tool] is not. I mean, either you get support or you don't, and on the back side we got a little support [...] 
``[The director] wondered what the algorithm would do if it were deployed at that point for [their family, when they were called into child services] and [their] assumption was that would not help [them] out, because [the director is] a minority, a teen [parent]. [In] reality, the algorithm probably would have really benefited her, because that would be considered an extremely low risk case.'' (L04) 
\end{quote}
%\khcomment{Hm... more broadly, I'm wondering whether the quote above adds value. It seems like it needs some more contextualization in order to be legible to readers (e.g., explaining what ``low risk case'' means in this context). But it's not clear the quote adds much insight (at least as currently written), compared with just omitting it and moving on more quickly to the next point.}
% ``Reality, the algorithm probably would have really benefited [them], because that would be considered an extremely low risk case, regardless of what the ailment was with this, with the son, and so that that has stuck with [them] [their] entire life and um, [they] now associate algorithms with something that's negative, just because [their] situation turned out positive in the day because [they] could talk to people. '' (L04/RD03) 
Many other leaders, like this participant, believed that opposition questioning the fundamental value of predictive AI tools was not well-founded. The overall sentiment that there are concerns to applying AI tools to certain contexts which are under-acknowledged by most agency decision-makers resonated especially with one R\&D Worker: 
\begin{quote} 
``I am much more conservative about the use of algorithms, especially with communities that tend to get the short end of the stick in terms of power and decision making [...] I think there's lots that can be done, but I'm not sure that decision making in these contexts is how I would have used algorithms. But that's just me.’’ (RD02) 
\end{quote} 
The participant had observed several instances of their colleagues developing AI tools, and through the years, had raised concerns around \textit{who} the AI tool was designed to benefit. %However, they described that most of their colleagues (e.g., other researchers, developers, analysts) overlooked their concerns. 
For example, during the design activity, the participant shared concerns around how the agency prioritized creating AI tools that helped achieve their own objectives (e.g., federal outcomes) rather than objectives that would benefit community members: 
\begin{quote} 
``I am here to tell you [when] these conversations have happened in agencies, the outcomes tend to be the things the agencies are interested in, not the things that the people that are being served by the agency are interested in. And so I think that's a real concern. People are tired of me saying, `This seems to be all about us--what works for us.'’’ (RD02) 
\end{quote} 

While uncommon across Agency Leaders and R\&D Workers, participants described that the above concerns were widely shared across frontline workers.
%\accomment{I'm having trouble squaring this with the above point that few participants voiced fundamental concerns. Is this because we didn't include enough frontline workers?}. 
In the next section, we discuss how participants interpreted and responded to internal concerns around the design and use of AI tools, which often came from frontline workers. 

% \khcomment{General note for the next subsection (and possibly elsewhere in the findings section: There are various spots where the text refers to ``participants''. But I think it would be really helpful to explicitly name which groups of participants are being talked about. On that note: I also had the sense while reading that this subsection presents leadership and developers' views about those closer to the frontline, without nuancing these views by presenting the perspectives of other groups. I was expecting to see more of this, at least by the end of the subsection. I think doing so would greatly improve the story, and provide readers with a clearer picture!}
% Anna note to self to address above comment later 

\subsubsection{\textbf{There are disconnects between frontline workers' AI-related concerns and their concerns as construed by agency workers with higher institutional power.}} \label{frontline_concerns}
% Alternative title: Frontline workers' concerns around AI tools are misconstrued by agency workers with more institutional power
Agency decision-makers believed that agency-internal contention around their use of AI were especially pervasive amongst frontline workers. In this section, we describe how agency decision-makers' interpretations of frontline workers' concerns shaped their goals for communicating with frontline workers on topics related to AI. Importantly, throughout our interviews, we observed discrepancies between the underlying causes for concern frontline workers voiced and the perceived causes attributed by agency leaders and developers. 

% THIS IS ABOUT RESPONSES TO CONCERN - ``As a result, agency decision-makers described a range of efforts to promote an organizational culture of understanding of AI, believing that disclosing how its agency would and would not use AI could help mitigate concerns.'' 
Agency Leaders, R\&D Workers, and Frontline Leaders often described efforts to ``convince'' frontline workers in their agency of the value AI-based approaches could bring, to help mitigate their concerns about AI-based tools. However, they often described that their  efforts to inform and align frontline workers' perspectives towards AI tools seemed futile, given the persistence of frontline worker concerns. For example, one Agency Leader who leads their department's efforts on developing new AI tools explained: 

\begin{quote}
``It's not priority [for frontline workers] to be using or [seeing] value in data-driven approaches like predictive analytics [...] It is a challenge [to] convince other teams that okay, by [...] using this approach you actually stand to gain [...] xyz aspects of things, you know. If your folks, if your staff gets read up sooner or your resources get [...] more efficiently used, then it is a win-win situation. There are challenges in trying to convince non-data people that there is merit to using a data driven approach.'' (L03) 
\end{quote}

Reflecting on their past experiences talking with frontline workers, another R\&D Worker with experience leading the development of new AI tools for their agency described: “It seems that no matter how much we voice that, that concern is present. [It's] hard [to] have people really grasp that the tool's meant to help inform their decision [...] rather than replace them as the decision maker'' (RD04). 
% \khcomment{Do all participants agree on this? Or do some participants have a different perspective regarding this sense of futility, and the persistence of frontine worker concerns?}

Beyond talking with frontline workers about the potential benefits of AI tools, some participants also believed that their AI evaluation approaches can help promote a more accurate understanding of AI across their agency. In particular, by applying model accuracy and fairness measures on human decisions, one participant hoped it would help reduce frontline workers' distrust towards AI tools by conveying similarities between human and AI decision-making. The participant, who worked in their agency's AI development team, described: 
%\khcomment{Thinking about the overall story/flow of this subsection: how does this point relate to prior points discussed in this subsection?}

\begin{quote}
``I think that one of the goals, at least in my mind, is to kind of change, the perception of what an algorithm is in like a lot of the kind of non-technical portion of the agency [...] % to where, because a lot of people who don't necessarily know the ins and outs of how machine learning works kind of, through popular media and stuff, learn that these algorithms can make mistakes that are that are very costly. And so 
they're very on board with the idea of algorithms and fairness, without realizing that the way that they're making decisions is kind of [...] an algorithm in and of itself.'' (RD05)
\end{quote}

% THIS IS ABOUT GAPS BETWEEN INTERPRETATIONS OF CONCERN AND ACTUAL CONCERNS -- However, talking with frontline workers who were involved in sharing feedback on piloted AI tools, we observed critical gaps between frontline workers' underlying concerns and the reasons for concern that agency leaders and developers attributed (Section~\ref{frontline_concerns}). 
Importantly, by talking with frontline workers who had experience piloting new AI tools, we learned that frontline workers were concerned about a broader set of challenges that come from AI use in social services, than what agency decision-makers we talked with had acknowledged. Agency leaders and R\&D workers often described the lack of technical literacy amongst frontline workers, often referring to them as ``non-technical'' portions of the agency, impeded their ability to seeing the value in AI tools. For example, during their interview, one R\&D worker described:

\begin{quote}
“The perception of what an algorithm is in a lot of the kind of non-technical portion of the agency [...] because a lot of people who don't necessarily know the ins and outs of how machine learning works kind of [learn] through [...] popular media and stuff [...] There's almost like a mystique around machine learning algorithms, like there's some amazing thing that is all knowing and all seeing, and therefore can predict all these different things.'' (RD04)
\end{quote}

Agency decision-makers also often believed that misconceptions on how their agency would use AI, rather than potential problems with the design of an AI tool, could explain why frontline workers were concerned. For instance, the R\&D Worker continued to describe frontline workers' fear towards AI as a problem of not understanding how AI will be used: 

\begin{quote}
``In general one of the things we do meet often [from frontline workers] [...] is the general distrust of algorithms and fear that it's going to replace people's jobs versus help them inform the decisions rather than take over the decision for them. [...] it seems that no matter how much we voice that, that concern is present.'' (RD04)
\end{quote}

% For example, reflecting on a past experience in which the prior director halted the deployment of an AI tool due to concerns around AI, \khedit{L04 said that}\khdelete{the participant described} they believed \khedit{it would have been valuable to clarify to}\khdelete{ communicating to} the director the intended role of the AI tool (to assist, not automate, decisions)\khdelete{ would have been valuable}. The participant continued to describe how, even when hiring directors of agencies, understanding the value of data-driven approaches to decision-making in human services is and should be a key consideration. The Agency Leader believed that focusing on ensuring this criteria is met could have prevented what they perceived as a setback in the agency's efforts towards improving human service decisions at the frontline. 
\begin{comment}
\begin{quote}
``I think data literacy would be something that everyone would benefit from. [...] If I had to redo it again, and I had some control over the process, I would have been part of the hiring committee and would ask the question about using research [...] that question needs to be asked of directors, I mean, [a prior leader] changed out a fair number of directors, because they were not data informed. And so then we wind up hiring one who I wouldn't say was anti, but [they] certainly wasn't pro use of data.'' (L04)
\end{quote}
\end{comment}

However, frontline workers we talked with emphasized concerns they had about how well agency decision-makers in their organization understood the impacts of AI tools in their work processes. They desired broader acknowledgement of the downstream consequences that deploying and maintaining an AI tool can have on frontline workers' responsibilities and labor. For example, during the design activity, one frontline worker who was involved in piloting their agency's new AI tool expressed frustration with claims they heard that the AI tool only requires workers to ``push a button'': 

\begin{quote}
``This tool is only going to be as good as the data. [...] How is that [going to] work? [It requires] someone that can actually monitor ensuring that the data that's being entered in is done consistently [...] What's that saying? Death by a 1,000 cuts like, you know, like folks are like. `Oh, it doesn't hurt that hard. It's not hard to push that button, or you know. Do this, or do you know, like what's so hard about it.' Well, folks that give that kind of feedback don't understand to kind of work that we do.'' (F02) 
\end{quote}
%\khcomment{Readers may need some help to understand what ``the branches'' means in the quote above.}
They described that, because each agency has different priorities for what data to input into their system, some model features may not be consistently documented in specific agencies. For state-level AI tools, frontline workers are responsible for carefully collecting and inputting data for the model. In one case, an R\&D Worker from another agency recalled a researcher flagging the added labor for frontline workers that would come with deploying their new AI tool. While there were suggestions to roll out the AI tool alongside other technical updates to help alleviate frontline workers' workload, there were infrastructural barriers to doing so. In particular, the proposed solution-developing a system that automatically populates data fields with needed information about a given case--could not actually be implemented by the development team, because they did not have maintenance control over the data inputting system. In the next section, we elaborate on how frontline workers also expressed a desire to inform key decisions on the design and use of AI tools that would directly impact their work.

% \hhcomment{We can consider highlighting the perceived role of media as another example of entities above the organization.} 

% “If a decision was made to move forward with with the analysis, then we would likely compute the audit. The historic quote, unquote algorithm experience of the status quo decision making by humans over the previous time frame of the data that we have available, and then could have conversations around the the implications of that, and perhaps that would inform if we were to ever develop an algorithm a machine learning algorithm for them, and considerations around which definition to choose and provide a baseline to to try and um improve upon” (RD04/RD05)

%\subsubsection{\textbf{Ad-hoc and limited avenues for domain experts (e.g., frontline workers) to inform AI design}} \label{frontline_feedback}
\subsubsection{\textbf{Frontline workers desire more collaborative and early-stage participation on AI design}} \label{frontline_desires}
% Make sure this narrative is clear: Moreover, while agency decision-makers described that they involved frontline workers in decision around AI design and use, frontline workers expressed that their existing pathways to inform decisions were too constrained; they desired opportunities to inform key decisions at earlier design stages (Section~\ref{frontline_desires}). 
%hhcomment{Higher order conclusion: the need for a standardize, formal, and collaboratively design process to engage with the frontline }
Participants across agencies described various touch points with frontline leaders and workers. In interviews with a couple of agencies, agency decision-makers described that Frontline Leaders were asked to check whether the most heavily weighted variables in a predictive model ``makes sense'' to assess its face validity. Agency decision-makers also described involving some Frontline Workers (those who are asked to use the AI tools they develop) in later stages of their development process, in particular, to pilot their tool. When we talked with the frontline workers who were involved in the piloting process, however, we learned there were challenges to having frontline workers share feedback on the AI tool. Aligning with the overall sentiment that there are major communication disjunctions between the frontline and higher ups who have more decision-making power over AI, one frontline worker shared that their peers were going through the motions of running the AI tool, but not actually critically interpreting or evaluating it. The worker hypothesized several reasons for this ranging from potential fear of insulting the AI developers, to not having enough time to notice and critically evaluate the model, to confusion around what feedback the AI developers were asking for. 

Frontline Leaders and Frontline Workers we talked with desired opportunities to more proactively engage with developers on discussions surrounding the design, evaluation, and use of AI tools. During the design activity, participants described that in current development processes, they did not have an opportunity to inform consequential decisions around the design of the model, such as what target the model should be trained to predict, as well as organizational polices governing model use, such as whether workers can exercise discretion in the use of the model. While frontline leadership were more likely to be consulted in middle (rather than later) stages of the AI development process, one Frontline Worker described that those in leadership positions may forget what the day-to-day work looks like on-the-ground (e.g., constantly ``putting out fires''), thus limiting their ability to anticipate downstream consequences from deploying models.

\subsection{Power Relations with those Above: Constant Pressures and (Lack of) Support from External Institutions}\label{above}
% Why this section is interesting + what the findings include.
Agency leaders and developers frequently described how powerful entities outside of the agency---e.g., federal government, state courts, and contracted private companies---shape their decisions around AI tools. Whereas prior empirical research has largely focused on how agency decisions and policies have downstream impacts on communities and frontline workers, our findings surface how legal and infrastructural barriers stemming from other institutions of power impact these decisions. For example, participants expressed frustration on their ability to interrogate ethical considerations on AI adoption decisions--a practice that had been largely ignored by contracting companies who hold a dominant negotiation position and are not incentivized to address the agency's concerns.

%In this section, we discuss how agency leaders and developers grappled with .... \akcomment{include findings list}. 
 
%\hzcomment{The role of department of justice: %\url{https://apnews.com/article/justice-scrutinizes-pittsburgh-child-welfare-ai-tool-4f61f45bfc3245fd2556e886c2da988b}}. \akcomment{consider adding another example, maybe creation of NIST framework}. 

%\hzcomment{This section + tensions within is less discussed in the literature overall. There is more literature on community interactions (from the perspective of community members). The community section complements this existing literature}

\subsubsection{\textbf{Agency decision-makers that procure AI technologies felt their relations with private companies fundamentally constrained their power on ensuring responsible AI design.}} \label{private_incentives}
Participants working in a city-level government agency that procured AI-based tools from private companies described a range of tensions around their interactions with these companies. In these settings, participants expressed frustration that private companies that develop their procured technologies often have minimal incentive to follow responsible and collaborative AI development processes.  %In contrast, participants at public sector agencies where AI tools were developed in-house by researchers, analysts, and developers employed by their agency seldom described similar challenges. %\khcomment{Possibly delete or reword the previous sentence? As it stands, it's not clear how agencies that develop tools in-house could face similar challenges (and given incentives, it's not so clear that we would have heard about these challenges, even if they do arise in-house)}

% Why there are power imbalances from procurement contracts that limited their ability to ask for things. 
Participants whose agency procured AI technologies from private companies described that they perceived power imbalances in their relationships with private companies. Participants described their relationships with private companies were often heavily shaped by the agreements listed in their procurement contracts with the company. Procurement contracts help define the conditions of the relationship between the city government department and a given private company that has agreed to develop a new technology. Participants described it can be difficult to anticipate what to include in the procurement contract before the AI tool is actually developed, leading to situations in which they have limited leverage in requesting certain model information and performance requirements. 

For example, one participant leading efforts to technologically modernize the region their agency serves described an experience learning about the importance of carefully wording procurement contracts, after being denied of important information from the private company. In particular, the participant shared an anecdote about how someone in the department noticed an AI-based risk assessment tool that evaluates road conditions was disproportionately flagging roads located in affluent neighborhoods. The department requested more information on this concern from the private company, but was denied any details: 
\begin{quote}
``One of our city Council members very quickly realized that the scores were prioritizing investments in areas that already are known to be some of our wealthiest neighborhoods. [...] Once we went back to the vendor to request [an explanation], the vendor said: `We cannot share that information with you because it's proprietary.'''(L05)
\end{quote}

% How there is limited incentive and leverage for improved development through procured contracts 
Agency stakeholders believed that their ability to use stringent procurement contracts was very restricted in practice. In particular, they expressed concerns that if the language in a procurement contract is too strong, private companies could simply decide not to sign the contract and instead go to another customer where the procurement contract is more favorable to them. The participant expressed their frustration around the lack of effective incentives for private companies, describing: 
\begin{quote}
    ``[T]here are fundamentally no incentives, rules, or regulations to incentivize the private sector to unlock or show how those systems work. [...] I have had companies tell me directly, point blank that it is not profitable to be ethical. And because of that, they really don't have anyone holding them to the fire to actually change their business practices [...] to reduce bias or achieve some of these goals around equity.'' (L05)
\end{quote}

% Challenges from following the SaaS model 
Participants at this agency also described new challenges surfacing from technologies that were procured from private companies before AI-based technologies were widespread. Because older procured technologies followed the SaaS (Software as a Service~\cite{turner2003turning}) model, the participants described that private companies often integrate AI-empowered features in their software updates without informing the technology users or city government. Without an awareness of which government technologies now use AI-powered features, participants described that workers may not be well-prepared to identify and account for possible technical failures from these updates. Therefore, one participant described that one of their projects includes documenting where AI is used across their departments. 

% Participants described how incentive misalignments and power imbalances \hzcomment{Might want to explain what the power imbalances are like here.} with private companies challenged their ability to deploy responsible algorithms through these processes. 

% Desires 
Participants also described an overall fear of over-relying on procured technologies from the private sector in the longer-term, given that this may lead to allowing private companies to have irreplaceable power over city departments. They described that it risks allowing the vendor to have ``a lock with the city,'' making the agency more susceptible to losing an understanding of the behind-the-scenes mechanisms around how their technologies are serving the public community. 

While they desired support for crafting more power-balancing procurement contracts, during the design activity, participants also acknowledged that any interventions at the procurement-level could only provide a limited amount of leverage to the city government (as described above). As a result, participants desired ways to incentivize private companies to prioritize ethical considerations in the design of their algorithms. For example, one participant described that having private company vendors be fully or partially liable to lawsuits resulting from the use of AI tools could encourage them to prioritize responsible development values. Participants also desired mechanisms to share experiences and information across public sector agencies, like enforcing a certification or ratings system where agencies could signal the quality of private company vendors they worked with in the past. To evaluate the effectiveness and impacts of new AI tools in a more collaborative and iterative approach, one participant also expressed an interest in exploring the use of digital test beds. 

%\begin{quote} ``It's understanding from those vendors. you know you can have your proprietary model and how you develop these algorithms. But without fully educating your customers on how you develop those algorithms and what safeguards that you have in place. It is doing a disservice to any of those customers, and especially those public sector customers like a city. who is trying to best protect the interest of our residents and still operate in the most efficient manner which some AI models could offer. But if there's not that communication in between, or a requirement to release certain information, then it makes it incredibly challenging. And we are at that, You know. center role where we've got to try to find the answers before any procurement is made, and that's difficult to do across 40 plus departments, all making their own decisions in a silo’’ (L05/L06)\end{quote}

\begin{comment}
    \begin{quote}
``We're thinking about like building test bed environments where we basically open up the city's digital architecture in a segmented network. So it's like protected from the rest of the city's network. And we invite private partners to come and test out a solution or technology in an environment where we require them to be transparent and open with us, we help them inform and improve their products. and then they can come out of a testing environment with a product that is [...] more ethical.’’ (L05/L06) 
\end{quote} 
\end{comment}

\subsubsection{\textbf{Agency decision-makers desired federal support that is better contextualized to their actual needs and challenges.}} \label{federal_support}
% Responsible AI guidelines are helpful in providing guidance and balancing incentives. 
Participants working in a city-level government agency also described that the federal government provides support by publishing guidelines that inform their efforts towards algorithm design and evaluation. For example, they described that the NIST AI Risk and Management Framework shaped their pre-deployment evaluation priorities for AI tools:
\begin{quote} 
``The Federal Government is behind cities. So cities are really at the front lines of innovating a lot of these tools, policies, and procedures, and the Federal Government, you know, often invite cities to consult on how they should develop national strategies or frameworks for this work [...] generally they push out information, guides, etc., that cities then adopt [...] because of that Federal backing of those [NIST cybersecurity] standards, the private sector has revised their approaches in order to be in alignment with NIST standards.'' (L05) 
\end{quote}

However, participants also described that federal guidelines often overlooked real-world resource constraints in their agency and government (e.g., by requiring creation of new roles). For instance, one participant described:
\begin{quote} 
``I find a lot of AI guidelines are very aspirational and not very realistic for municipal government, because they don't address the capacity issue and because they require in many cases the creation of new types of roles and positions within the city that don't exist in many cases. So, for example, an algorithmic auditing role: someone whose purpose within local government is to process and evaluate AI tools for their abilities to meet certain standards that also don't exist yet.'' (L05) 
\end{quote} 

Participants further imagined that federal guidelines could help align incentives towards responsible AI development between the city government and private companies. One participant explained that, while city government workers themselves have limited leverage, standards that are ``backed'' by the federal government are more likely to nudge private companies to adjust their development approaches. However, the participant described that federal responsible AI guidelines alone may not be enough to improve private sector companies' practices. They expressed concerns that., without proper ``incentives and enforcement,'' guidelines could be ignored by private companies: ``If you're setting a guideline for, you know, private sector companies, how realistic is it given that they're really, main incentive is the bottom line, and to remain solvent as a company'' (L05). Therefore, the participant advocated for federal requirements to mitigate the risk of private companies simply rerouting their services to other (less ethics-prioritizing) customers.

\subsubsection{\textbf{Agencies grapple with concerns about AI outputs being misused in state courts.}}\label{court_risk}%\accomment{may want to change to "grapple with concerns"}\label{court_risk}
Several participants raised concerns around how risk scores generated from AI tools could be used outside of their agency, in legal cases at state courts. %Participants were concerned about whether and how these these AI risk scores could be misused or misinterpreted by attorneys in legal cases. Acknowledging that public sector services frequently become involved in legal processes, they desired ways to more proactively interact with attorneys and legal experts who were at the forefront of these engagements. 
% Concern of misuse in courts, by attorneys 
Participants expressed concerns that AI risk scores could be misused or misinterpreted by attorneys in the court. For example, frontline workers imagined that attorneys might use ``incorrect'' risk scores to their advantage. One frontline worker, who piloted an AI tool designed to assist reunification decisions for children placed out of their home, recalled feeling concerned that an attorney would overlook the workers' judgements in favor of the AI tool's risk score:
\begin{quote} 
``... but we were predicting that attorneys are going to say, look, [the agency] has this tool [...] that says this family would probably be just fine if they return these kids home. And meanwhile we're over there saying there's no way this family is ready. They've done nothing. They're using substances currently or [...] There's violence in the home.'' (F03) 
\end{quote} 

Some leaders and developers also described experiences receiving advice from attorneys to avoid developing a new AI-based decision-making tool, to mitigate the potential for producing additional information that might be used against them in a lawsuit. One participant described that, in existing AI-based decision-making tools, they attempt to mitigate the risk of score misinterpretation by including a text description indicating what a risk score does and does not indicate. However, participants acknowledged the limited leverage they currently have in preventing legal misuse of AI-based risk scores. 
\begin{comment}
    \begin{quote} 
``[I]t caused me to even say at 1 point I'm not even gonna run this until I'm already in the process of sending kids home. Because then I don't have to deal with an attorney asking me why I'm not sending kids home.’’ (F03)
\end{quote} 
\end{comment}
\
\subsubsection{\textbf{Some agency decision-makers---but not others---want to ensure their decisions about AI are better informed by legal considerations.}} \label{policy_desire}
% Desires for earlier discussions with attorneys and legal experts 
To both inform the design of AI-based tools and inform legal experts of the processes and capabilities of AI-based tools, several participants expressed during the design activity that they desired earlier discussions about the model with legal experts. Describing the fundamentally intertwined relationship between child welfare screening and the legal system, one frontline worker %(and several other developers \akcomment{consider including participant numbers}) 
emphasized the importance of considering legal perspectives early and often: ``A child welfare is based on policy, right? It's based on laws. There's no way around it'' (F02). Another participant further elaborated: ``Policy, procedure, statute rules [...] all need to be taken into consideration [...]. You need to be having conversations with your Department of Justice and attorneys'' (RD06). 
%Some developers imagined that legal experts could provide valuable insights on how to appropriately mitigate algorithmic fairness concerns in AI tools or ensure that those tools work within legally defined bounds. 

Participants described that their cognizance of how legal considerations could shape the use of AI-based tools was based off of prior experiences they had interacting with the legal system due to the design of their AI tools. Participant described that these interactions often occurred in a reactive, ad-hoc manner, for example, through lawsuits against the agency: ``In our case there was also a lawsuit that was filed, which changed the way we responded to the different needs of the system'' (L03). This participant imagined that, by working in collaboration with legal experts who were familiar with existing statutes and laws, they could help bridge new understanding of how these considerations play out in AI development.  

However, not all participants were interested in having collaborative discussions with attorneys and legal experts. For example, one agency leader believed that attorneys were overly ``conservative'' towards the use of new AI-based tools:
%\begin{quote}``And so [the attorneys] started, going out on many more cases than what they had been previously. And so the algorithm wasn't used because, uh, they were looking at uh a legal definition of what to go out on and what not to go out on. And there's significant subjectivity in in that.’’ (L04/RD03)\end{quote} 
\begin{quote} 
``Well, I try and stay away from the attorneys for a lot of reasons. They're always on the Conservative side, and there's no consideration for the benefit to the family. They only consider the risks and not the benefits.'' (L04)
\end{quote}
This participant, along with another developer, was concerned that policy and legal considerations compromise the potential usefulness of developed AI tools, hampering innovation in the longer-term. In the next section, we discuss participants' perceptions of and attitudes towards engaging with other stakeholders to their deployed AI tools--particularly, those situated ``below'' the agency.
%Interestingly, some participants also expressed interest in opportunities to inform the design of new \textit{policies} that guide the development or use of AI tools: 
% \subsubsection{\textbf{Frustrations and tensions with law and policy actors}}
% Tensions and tradeoffs with considering legal perspectives in the design of AI 

\begin{comment}
Another participant expressed frustration with the lack of clear guidelines for compliance.
\begin{quote} 
``There is no sort of central policy that [...] is adopted across the departments, or that the departments are all aware of. There is no culture [...] around automated decision making that it influences how our city departments make decisions [...].’’ (L05/L06) 
\end{quote}     

They thought this problem is compounded by the slow rate at which policy changes happen at the Federal level:
\begin{quote} ``[On the policy side] not a lot was getting done. A lot of the work that I was involved in from a policy perspective was just became repetitive year over year, because Congress wasn't getting anything done. I moved around the time the government shut down [...] very divisive political environment at the Federal level.’’ (L05/L06) 
\end{quote} 
And the frequency of leadership changes at the state level:
\begin{quote}
    ``Most states are affected by governor changes and by director changes, not by lawsuits.''(L04/RD03) 
\end{quote}
%%%%%%%%%%%%%%%%%%%%%%%%
\end{comment}

\subsection{Power Relations with those Below: Barriers to Hearing and Empowering Impacted Communities} \label{below}
Recent research has increasingly called for community participation around the design of public sector algorithms. As a result, the participatory machine learning and human-computer interaction communities have proposed a range of methods for collaborative AI design with diverse stakeholders, including those without technical expertise (e.g.,~\cite{kuo2023understanding}). However, we know little about whether and how these design approaches and tools are actually used in practice. Furthermore, developing an understanding of the real-world factors that shape existing agency-community collaboration around AI development processes is critical to ensuring that the participatory model and design approaches can actually be transferred and used in practice.
% Moreover, while some prior work has explored community members’ concerns around the use of AI tools in the public sector, this work often leaves open questions around whether and to what extent these concerns have reached those who have the power to shape external collaborations around AI development processes.  
% \akcomment{make motivation more precise, e.g., participatory ML work has proposed different approaches, this study surfaces additional considerations (almost prerequisites) to having these approaches used in practice. }
% \akcomment{include above, some of our interpretation. no incentives - lack of progress on building infrastructure, overcoming communication barriers, etc. upfront that this is a root cause. But also downstream effects. } 

In this section, we shed light onto infrastructural and social barriers that are often overlooked when designing approaches for community participation around pubic sector AI development. Overall, we found that individual agency leaders and R\&D workers expressed interest in involving community members (Section~\ref{community_interest}) but, without formal incentives to set up these collaborations, they faced downstream challenges when attempting to advance community engagement efforts in practice. These challenges further disincentivized agencies from expanding participation around their AI development pipeline. For example, participants described a lack of infrastructural support in identifying and connecting with relevant groups (Section~\ref{community_participation}). Moreover, some agency leadership and developers were cognizant of power and knowledge differentials between the agency and community members, which they believed would challenge R\&D workers’ abilities to have collaborative discussions with community members around AI development (Section~\ref{community_communication}). 
 
%We highlight how %, despite overall sentiments from the HCI research community that public sector agencies may \akcomment{change this word maybe: devalue} community participation around AI~\cite{}, 
%several participants with decision-making power at public sector agencies expressed concerns around their lack of community engagement and desired pathways for supporting more participatory AI development processes. %\khcomment{I'm not sure the first part of this sentence accurately captures overall sentiments from researchers in HCI, participatory ML, etc - can reframe around what sorts of methods research in this space tends to contribute (and what challenges they do / don't address)?. This could help in making the claim here more precise: clarifying what we think many in the community might find surprising!} 

\subsubsection{\textbf{Agency decision-makers expressed interest and value in involving  community members in decisions about AI design.}} \label{community_interest}
Some participants (RD02, RD04, RD07) expressed interest in expanding their agency's currently insulated approach to developing AI tools by involving community members in their development process. For example, during the design activity, one participant (RD02) acknowledged that leaders and developers internal to their agency likely view, interpret, and make design decisions around the AI tool differently than those ``outside'' of their agency, like community members. 
\begin{comment}
\begin{quote} 
``But \khdelete{but }it \khedit{[...] }\khdelete{had to do with It }was related to the people in different seats at the table. Right? Yeah, the same question. The same decision might be viewed differently by people sitting at different seats.’’ (RD02) 
\end{quote} 

\khcomment{It's not clear that the quote above is adding much insight. So I could see deleting this quote, to help make the section more focused!}
\end{comment}
Another participant expressed frustration that their agency appeared to develop AI tools that prioritized the agency's needs over those of the community, e.g. developing AI tools to meet federal targets that may not be aligned with community needs. 

% \hzcomment{I deleted the quotes about how behavioral science/domain knowledge is ignored in the design process. I think there is a difference between behavioral science knowledge and community needs, and we focus on the latter in this section. }
% \hzdelete{The participant further described how they felt their concerns were sometimes dismissed or ignored by their researcher peers. They recalled an experience where a researcher analyzing administrative data had grouped ``reunification'' with ``adoption'' and ``guardianship.'' When the participant pointed out that those terms were meaningfully distinct in the social welfare system, the researcher had accused them of ``not understanding their model’’ \akcomment{find quote} (RD02). }
This participant further elaborated that, if public sector agencies do not begin listening to community concerns and collaborating with communities around the tools they build, then they are likely to miss opportunities to address fundamental, root problems and risk focusing a disproportionate amount of attention towards fine-tuning algorithm designs. The participant advocated for the importance of engaging community members, arguing that only by meaningfully engaging with those ``closest to the problem'' can agency researchers move past their own privilege and blindspots. Without engagement with communities to understand their true problems, researchers ``can't get out of our own way'' (RD02). 
% \akcomment{NOTE -- consider splitting developers/researchers into --> scientists and managers? or just scientists and developers? look into role names} \hzcomment{I think we might not want to introduce new role splitting at the end of the findings.} \hzcomment{I deleted a few quotes here. I think they are long and do not provide additional new insights. }

% Sees the value – limited knowledge, need to be able to integrate diverse perspectives. 

% One participant was especially concerned that leaders and developers have limited perspectives and knowledge specifically around \akcomment{include details, something about downstream impacts on community members, domain-specific knowledge, how workers use the tool, etc.}. 

\subsubsection{\textbf{Agencies are still building out infrastructural support to sustain community partnerships around AI}} \label{community_participation}
While participants were interested in improving their agency’s practices towards involving community members in AI development, they felt there was inadequate infrastructural support to do this in practice.% (\hzedit{participant nums})
%Other participants% (\hzedit{participant nums}) also anticipated that communication barriers, from community members’ lack of technical knowledge, would hinder their ability to meaningfully engage community members. 
Other agency leaders and developers %(\hzedit{participant nums})
anticipated potential power and cost imbalances between the agency and community, and desired ways to mitigate potential harms arising from such concerns. 

% practices from agency 1 
Participants at different organizations had varying levels of structure in their practices for involving community members in their AI development process. At the agencies where algorithms were developed in-house, one agency described during the interview that they received feedback on ethical considerations around developed AI tools from an external research committee, which involved parent advocates and people who have lived experience in foster care. 

% Practices from agency 2 
On the other hand, participants from another agency that develops in-house AI tools  described facing several barriers to involving community members. Reflecting on how they want their development process to improve in the future, during the design activity, participants expressed interest in receiving feedback on their AI tools from community members; however, they described that their agency does not currently have the infrastructural systems in place to support and sustain community engagement around AI development. To learn how to establish pathways for involving community members, some participants described plans to talk with other departments in their agency (e.g., an office of equity). Participants (RD06, RD07) described that the agencies need guidance on identifying community members to reach out to, ensuring that participation is accessible, determining logistics around payment and involvement time, and establishing partnerships with mutual trust: ``It's just a different piece of the puzzle here to get that input’’ (RD06). The participant, who manages a team responsible for implementing newly developed AI tools into specific decision contexts, further elaborated: 

\begin{quote} 
```It's building that network of people who want to come in and give feedback and participate in these [activities]... If we really want genuine input and people to be coming in from external to our agency, we need to make these things more accessible and more available to them. They're not during like eight to five business hours Monday to Friday. We [should] do groups after five on weekends. Can we pay people for their involvement? These are conversations that we're having right now. So that it's not some false inclusion and access conversation. It's like true participation.'' (RD06)
\end{quote} 

In addition to navigating infrastructural challenges around following appropriate logistical norms (e.g., payment, timing of sessions), participants were additionally concerned about their ability to navigate power and knowledge differentials between them and impacted communities. We elaborate on these concerns in the next subsection.
% \khcomment{Could potentially keep this next point brief, since it has been fairly well covered in prior literature. Or otherwise, make sure to highlight what is newer / more surprising here!}
% One participant further anticipated that receiving trust from community members may be challenging, given decades of history … : \akcomment{insert quote}. 

\subsubsection{\textbf{Agency decision-makers felt ill-equipped to effectively communicate with impacted communities due to power and knowledge differentials.}} \label{community_communication}
% Anticipated challenges with better engaging with frontline workers 
Beyond having more infrastructural support and guidance on building sustainable partnerships with community members, some participants also anticipated communication challenges with effectively integrating perspectives from those beyond their usual research and leadership team. 
For example, during the interview, one participant elaborated that well-intended developers may struggle to understand community concerns due to lack of lived experience, while community members may not have the technical knowledge needed to inform algorithm design. This participant recalled they have often seen ``perfectly well-intentioned analysts not hearing or understanding [local community members'] concerns,’’ even when agencies make an effort to engage with local communities (RD02). During the design activity, as the participant emphasized the importance of developing AI tools to meet outcomes that center community (rather than agency) needs, the participant expressed further concern that some of their peers may not take feedback from ``non-technical’’ people seriously. They described that they have always expressed concerns about the agency's siloed approach to designing algorithms but felt that the agency as a whole did not prioritize addressing these concerns (``[colleagues are] tired of me saying [this]'').  Another participant raised concerns about the feasibility of involving people without domain knowledge on AI in ways that bring value to the conversation: 

\begin{quote} ``I've seen a lot of the time that people who have no technical background, they may sit in this meeting right and just not have any idea of what the conversation is. When bringing everybody together, how do we communicate that both people with technical backgrounds and those with no technical background are able to follow the discussion on the conversation? I think that's hard. I think it's almost like the power differential to and from the technical folks.'' (L04) 
\end{quote}
% \khcomment{Should the last couple words above read ``technical folks''? (I'll admit that at some level, I'm kind of hoping this is not a typo, and there is actually a technical fox involved... in which case, I want to know more!)}

% \akcomment{code – Participant anticipates that a challenge of adopting the deliberation protocol would be ensuring there are no communication barriers between those with technical and non-technical backgrounds. They make an analogy to how this creates power differentials}

% \hzcomment{I deleted the "ethics" quotes and discussion. I think they are not very relevant to the themes of this section. }
A couple participants additionally acknowledged the power imbalances that might hinder effective collaboration between AI developers and community members, as they envisioned potential future challenges during the design activity. One participant anticipated that community members would have to endure higher emotional burdens from participating in AI design compared to public sector agency workers, as they would be ``asking a lot more of the communities we serve to come and dwell on those spaces than is asking of us'' (RD07). %\accomment{Should this be "than is asking of us"?}
They continued to advocate that government agency employees should take the initiatives to share the power and spaces, in order to remedy ``perceived harms from governments on the people’’ (RD07). However, participants (including RD07) described that their agency doesn't currently have the cultural norms and guidance needed to support effective power sharing that would be required in collaborative design processes that go beyond consultation:% \accomment{Is there a reason we use sharing power vs power sharing? I think it would be clearer to use the latter} 
\begin{quote}
``In government, to really honor [power sharing] and just not be like, you know, mining people for their opinions, but to really allow community to kind of drive some of these decisions a little bit, and determine the shape of predictive models and whatnot that sit on decision points that are important to their lives. That still feels a little foreign, and we--as in the government agency--I don't know that we have great mechanisms yet for some of that kind of power sharing.'' (RD07) 
\end{quote}
Overall, while participants acknowledged the need to involve impacted community members in their AI development process, in many cases, they felt that their agency was not well-equipped to actualize this need. We discuss implications for addressing these challenges--including the lack of infrastructural and social support in engaging stakeholders--in the Discussion section.

\section{Discussion}\label{sec:discussion}
 The SIGCHI research community has advanced several methods and tools to support the responsible development of AI tools. One particularly promising form involves expanding participation--from the earliest stages of AI design to after the AI model is deployed--to broaden \textit{whose} perspectives inform AI design and use decisions~\cite{kulynych2020participatory,wolf2018changing, delgado2021stakeholder}. In examining how agency decision-makers currently make AI-related decisions, our study strongly suggests that participatory AI methods and tools provide a particularly promising pathway to improving the responsible development of AI tools in public sector contexts. However, the research community's exploration of these methods and tools in \textit{real-world} organizations, especially in public sector settings, remains nascent. While hurdles to bringing HCI approaches from the research community into real-world practice are expected, identifying ways to streamline this process into public sector contexts remains a relatively under-explored and--as AI tools continue to proliferate across public sector settings~\cite{kawakami2024situate,loi2021towards,kuziemski2020ai,levy2021algorithms}--increasingly time-critical challenge. 
 % As public sector agencies rapidly create and use AI systems in high-stakes decision-making domains like social services, it becomes increasingly critical to consider \textit{who} or \textit{what} inform these decisions, and \textit{why}. \khcomment{to-do: see whether we can re-write previous sentence to improve clarity and simplify. As currently written, it may not be entirely clear what this means (in particular, the second part of the sentence: "well-informed, by reflecting on ..."} \khcomment{Need to include some sentences here to transition from the findings shared above to the specific focus on participatory approaches (below)} 

In situating public sector agencies’ current practices and challenges in a network of power relations, we surface a broader set of considerations--including regulatory, infrastructural, and cultural factors--that play a crucial role in the real-world adoption and implementation of participatory AI approaches. 
% \khcomment{Does ``research-based design methods'' refer to participatory AI practices? If so: would it make sense to say this directly (aligning language with that used above)?}\hzcomment{I wonder if we need a couple of sentences to motivate participatory AI approaches here, before we discuss the uniqueness of participatory AI in public sectors. Or we should broadly say "design and use of public sector AI", instead of jumping into participatory AI so quickly.}\khcomment{+1 Or possibly even earlier than this paragraph? i.e., in the paragraph above where participatory AI is first mentioned?}
Our findings point to unique implications for how to support participatory AI in \textit{public sector} contexts, and more broadly, suggests opportunities for future research to distinguish between public versus private sector needs and applications around implementing participatory approaches. In the following, we discuss several of these implications (Section~\ref{sec:agenda}) along with limitations and reflections on our research (Section~\ref{sec:limitations}).

\subsection{Power-Conscious Implications for Designing Public Sector AI} \label{sec:agenda}
Building on our findings, we propose power-conscious recommendations for researchers and policymakers working towards improving the design and use of public sector AI tools. By ``power-conscious,’’ we refer to the practice of centering networks of power relations–-including those that exist within agencies, between agencies and other institutions of power, and between agencies and the communities they serve---to identify how they could impact agency decisions around AI, and what concrete avenues for improvement exist. We note that there are both opportunities for future work to better \textit{mitigate} the negative effects of power relations, as well as to \textit{leverage} existing power relations to support more responsible practices. Towards realizing these opportunities, we discuss implications for mitigating and leveraging the network of power relations public sector agencies reside in, to support more participatory AI design, evaluation, and governance approaches. We first discuss research implications to reshape power relations with those ``within’’ and ``below’’ public sector agencies (Section~\ref{sec:agenda_1}). Next, we discuss policy implications to leverage existing power relations from ``above’’ to support this reshaping (Section~\ref{sec:agenda_2}).   
% \khcomment{Suggested small edits to avoid saying ``radical'' reshaping, since while the directions described below are important and impactful, they may be viewed as more pragmatic, incremental steps --- not necessarily the most \textit{radical} ways to fundamentally reshape power dynamics.}

\subsubsection{\textbf{Reshaping Power Relations With Those Within and Below}}\label{sec:agenda_1}
How can the research community better support public sector agencies in expanding participation around AI design, evaluation, and governance? Our findings point to opportunities to help refocus and redirect existing research efforts around public sector AI. In the following, we discuss research and design implications to support public sector agencies in reshaping power dynamics in interactions amongst those within (from agency leaders to frontline workers) and with those below (impacted communities). 
% \akcomment{make these bullets instead?} \khcomment{Yes, I could see making the below bullets!}
\begin{itemize}
    \item \textbf{Make visible the value of participatory approaches to AI design and evaluation to public sector agency leaders and developers}, for example, through case studies that showcase tangible benefits and metrics (whether quantitative or qualitative) that can clearly communicate the value of participatory approaches. This should \textbf{increase incentives for agency leaders and developers to explore and uptake participatory approaches}, while also supporting agency workers already doing the work of advocating for increased community and worker engagement (Section~\ref{community_interest}). Future research could identify metrics and outcomes that agency leaders find motivating, that additionally demonstrate the benefits of using participatory approaches (e.g., public perception, worker happiness and satisfaction). 
    \item \textbf{Ensure agency leaders and developers are aware of existing easily adoptable methods}, by improving the dissemination and findability of participatory methods and tools developed in the research community. Increasing awareness of the vast landscape of existing approaches can help lower agency barriers to exploring their use, by increasing the chances that agency leaders find an approach that meets their context-specific needs. In turn, awareness of the landscape of participatory methods may also help to mitigate misconceptions around \textit{whose expertise} is valuable across stages of the AI development cycle and the practicality of realizing this value. For example, agency leaders were concerned about the feasibility of involving impacted communities without domain knowledge on AI in AI development processes (Section~\ref{community_communication}) and believed frontline workers’ involvement should be consolidated to later rather than earlier stages of AI development (Section~\ref{frontline_desires}). In this case, ensuring public sector agencies are aware of approaches for effectively involving non-AI domain experts in early stages of AI design has potential to help mitigate these concerns.
    \item \textbf{Train agency leaders and managers on ways to effectively work with frontline workers and community members}, for example, by creating safe communication spaces for frontline workers and developing the reflexive skills needed to interpret and respond to worker concerns. While prior academic literature and regulatory standards currently focus on training for \textit{frontline workers} using AI tools, our findings suggest the additional need for training interventions at the leadership and managerial levels that could help foster an organizational culture of deliberate and safe cross-role collaboration. For example, the training could teach leadership and managerial workers about different forms of participation (e.g., from the Ladder of Participation~\cite{arnstein1969ladder}), to help caution against pitfalls like  ``participation-washing'' ~\cite{sloane2022participation} and scaffold ideation around new policies or procedures that could support meaningful cross-stakeholder participation in their agency. Training agency leaders and managers to create the organizational and social conditions needed for effective collaboration with frontline workers has potential, over time, to help reshape lopsided power relations between agency leaders and frontline workers. 
    % \khcomment{In the previous sentence, it may not be obvious to readers how creating conditions for effective collaboration between *leadership* and *frontline workers* this could help reshape lopsided power relations with *community members*. Perhaps delete ``community members'', or add more explanation here?}
    This can open more opportunities to fully understand (and avoid misinterpreting) frontline workers’ underlying concerns around AI (Section~\ref{frontline_concerns}). Moreover, learning these skills can additionally help address agency leaders’ doubts around their ability to effectively power share with impacted communities during design processes (Section~\ref{community_communication}).
    \item \textbf{Innovate on participatory AI methods and tools to better address problems of practice}. Develop methods and toolkits that provide \textbf{concrete guidance and support for \textit{social and political} tasks required}. For instance, these tasks might involve developing a deep, grounded understanding of the historical, social, and political context of the intended AI deployment setting (e.g., to help interpret AI evaluation measures, understand implications of data limitations, identify who is the ``stakeholder'' of the AI system versus who is a proxy to the stakeholder, amongst other tasks). Responsible AI methods and toolkits today primarily provide guidance on completing technical tasks, implicitly framing the work of AI ethics as one that is mostly technical~\cite{wong2023seeing}. Our findings suggest that, if responsible AI methods and toolkits continue to primarily provide support for work that is framed as ``technical'', \textbf{there is a risk of further reinforcing agency leaders and developers’ misconceptions around the roles and expertise that are well-equipped and necessary for doing responsible AI work} (Section~\ref{frontline_concerns}, Section~\ref{frontline_desires}, Section~\ref{community_communication}). 
\end{itemize}

\subsubsection{\textbf{Leveraging Power Relations With Those Above}}\label{sec:agenda_2}
% Next section: But agencies + researchers can't do it on their own. We'd also benefit from support from above. Here's what some of that support could look like!
Many of the implications discussed in the prior section can best be realized through additional policy-level interventions. In this section, we discuss policy implications to support the research community and public sector agencies in pursuing the opportunities described in Section~\ref{sec:agenda_1}. 
\begin{itemize}
    \item \textbf{Situate regulatory standards and guidance in public sector agencies’ on-the-ground needs and practices}, for example, by streamlining communication across different levels of government  around the personnel and resources currently available versus needed to do responsible AI work. Participants in our study described that their agency currently struggles to adopt ``best practices’’ proposed by the government (e.g., NIST RMF), given adoption requires the creation of new roles and additional resources (Section~\ref{federal_support}). While existing federal efforts to support the integration of AI into government services (e.g., AI Factsheet~\cite{house2023fact}) communicates the importance of providing guidance on AI adoption, our findings emphasize the importance of equipping agencies with the resources needed to follow that guidance. Mitigating resource gaps would require having \textbf{mechanisms for public agencies to communicate the (in)feasibility of implementing guidelines and, ideally, to request additional resources or funds}. For example, this can be supported by creating a technology unit in federal government (e.g., similar to the United States Digital Service) dedicated to supporting consultation and collaboration between federal and local agencies. This type of government program can help intermittently test regulatory standards in collaboration with local agencies, to experimentally implement standards and iteratively learn opportunities for mitigating resource-level challenges.
    \item \textbf{Provide government funding for research priorities that innovate new participatory approaches, including through examining the \textit{integration} of these approaches into specific organizational settings (e.g., public sector agencies)}. Few research projects on AI design innovation publicly demonstrate the use and integration of their proposed methods in real-world public sector agency settings. Our findings suggest there are several overlooked problem spaces at the intersection of developing and actually implementing participatory AI design approaches that work at scale across public sector agencies (e.g., around supporting the findability and adoptability of design approaches). Using government funding to incentivize and support research that demonstrates real-world integration and use of participatory AI design approaches can further help lower the barrier for agencies to adopt these approaches in practice.
    \item \textbf{Support public sector agencies in both learning and sharing with others their experiences implementing participatory AI approaches}, for example, through creating a national repository documenting case studies of integrating participatory approaches to AI design into actual public sector agencies. These guides can help public sector agencies more easily learn from each others’ experiences around translating research-based design approaches into practice. While participants in our study shared common challenges around developing and procuring AI technologies in each of their agencies, several expressed that they knew little about other agencies' experiences around AI deployment; as a result, they desired knowledge-sharing mechanisms to learn from other agencies' practices (Section~\ref{private_incentives}).
\end{itemize}

\subsection{Limitations and Reflections}\label{sec:limitations}
% Note - this is largely same as before 
In this paper, our goal was to understand how U.S. public sector agencies, particularly those providing social services, make decisions around the design and use of AI systems. To meet this goal, we interviewed participants working in public sector social service departments in the United States. Given differences in existing regulatory requirements, cultural norms, and other country-specific factors that impact AI design and development practices, our findings are more likely to generalize to similar agencies within the United States, not necessarily to those in other countries (as evident by findings from~\cite{veale2018fairness}). We also acknowledge that public sector agencies within the United States may differ in size, access to resources, compliance requirements, leadership inclination towards investing in AI systems, and many additional factors. Based on the authors’ informal conversations with other public sector agency leaders across the United States and prior literature, we believe that the agencies we studied may be relatively advanced in their thinking about AI applications and responsible development practices. Therefore, it is possible that the challenges discussed in this paper represent ones that public sector agencies encounter even when they are motivated to prioritize responsible AI practices. Moreover, given the nature of challenges described in the paper (focusing on infrastructural, legal, and social challenges), and the generality of their sources across departments (e.g., being under-resourced, having historically tense relations with the community), we believe it is possible that other U.S. public sector agencies may encounter similar barriers and challenges as those described in the paper. 

The act of \textit{studying up} public sector AI involves understanding barriers and challenges from the perspectives of those who hold significant power and responsibility over their existence. To help scaffold relevant reflections and deliberations around this potentially sensitive topic, our study included a design activity that asked participants to think about a hypothetical scenario considering the use of AI and asked them to describe their past experiences when discussing their responses. However, it is possible that participants were not comfortable being fully transparent about their experiences and challenges. Especially for practices and challenges that involve other stakeholders, a critical line of future work is triangulating observations across a broader range of relevant stakeholders, including frontline workers, impacted communities, and legal experts. %\khcomment{This last sentence seems out of place here. The previous sentence talked about participants' level of comfort in being fully transparent. But as currently written, this seems like a different point, related to the range of stakeholders we worked with in this study?}

Finally, agency decision-making is often made ``public'' through online documentation (e.g.,~\cite{goldhaber2019impact}) but a lot is still excluded from those documents. Understanding the latent perceptions, assumptions, and experiences underlying agency decision-making is crucial to improving development practices in the public sector. We are deeply thankful to the participants and their agencies for sharing their time and experiences with us, and hope that collaborations between public sector agencies and independent researchers (such as those that made this study possible) become more commonplace. 

% \khcomment{As a general rule, try to avoid ending a paper on limitations. Consider adding a very brief Conclusion section at the end!}

\begin{acks}
We would like to thank the study participants again for generously sharing their time, experiences, and thoughtful reflections with us. Thank you to our paper reviewers for their insightful feedback that helped improved this work. We also gratefully acknowledge the support of the Digital Transformation and Innovation Center at Carnegie Mellon University sponsored by PwC. This research was also supported by funding from the UL Research Institutes (through the Center for Advancing Safety of Machine Intelligence), the National Science Foundation (NSF) (Award No. 1952085, IIS2040929, and IIS2229881), the NSF Graduate Research Fellowship Program, and the K\&L Gates Presidential Fellowship (through Carnegie Mellon University). Any opinions, findings, conclusions, or recommendations expressed in this material are those of the authors and do not reflect the views of NSF or other funding agencies.
\end{acks}

\bibliographystyle{ACM-Reference-Format}
\bibliography{justiteam}


\begin{thebibliography}{60}

%%% the Web address.

\ifx \showCODEN    \undefined \def \showCODEN     #1{\unskip}     \fi
\ifx \showDOI      \undefined \def \showDOI       #1{#1}\fi
\ifx \showISBNx    \undefined \def \showISBNx     #1{\unskip}     \fi
\ifx \showISBNxiii \undefined \def \showISBNxiii  #1{\unskip}     \fi
\ifx \showISSN     \undefined \def \showISSN      #1{\unskip}     \fi
\ifx \showLCCN     \undefined \def \showLCCN      #1{\unskip}     \fi
\ifx \shownote     \undefined \def \shownote      #1{#1}          \fi
\ifx \showarticletitle \undefined \def \showarticletitle #1{#1}   \fi
\ifx \showURL      \undefined \def \showURL       {\relax}        \fi
% invisible to TeX
\providecommand\bibfield[2]{#2}
\providecommand\bibinfo[2]{#2}
\providecommand\natexlab[1]{#1}
\providecommand\showeprint[2][]{arXiv:#2}

\bibitem[Arnstein(1969)]%
        {arnstein1969ladder}
\bibfield{author}{\bibinfo{person}{Sherry~R Arnstein}.}
  \bibinfo{year}{1969}\natexlab{}.
\newblock \showarticletitle{A ladder of citizen participation}.
\newblock \bibinfo{journal}{\emph{Journal of the American Institute of
  planners}} \bibinfo{volume}{35}, \bibinfo{number}{4} (\bibinfo{year}{1969}),
  \bibinfo{pages}{216--224}.
\newblock


\bibitem[Bansal et~al\mbox{.}(2019)]%
        {bansal2019beyond}
\bibfield{author}{\bibinfo{person}{Gagan Bansal}, \bibinfo{person}{Besmira
  Nushi}, \bibinfo{person}{Ece Kamar}, \bibinfo{person}{Walter~S Lasecki},
  \bibinfo{person}{Daniel~S Weld}, {and} \bibinfo{person}{Eric Horvitz}.}
  \bibinfo{year}{2019}\natexlab{}.
\newblock \showarticletitle{Beyond accuracy: The role of mental models in
  human-AI team performance}. In \bibinfo{booktitle}{\emph{Proceedings of the
  AAAI Conference on Human Computation and Crowdsourcing}},
  Vol.~\bibinfo{volume}{7}. \bibinfo{pages}{2--11}.
\newblock


\bibitem[Barabas et~al\mbox{.}(2020)]%
        {barabas2020studying}
\bibfield{author}{\bibinfo{person}{Chelsea Barabas}, \bibinfo{person}{Colin
  Doyle}, \bibinfo{person}{JB Rubinovitz}, {and} \bibinfo{person}{Karthik
  Dinakar}.} \bibinfo{year}{2020}\natexlab{}.
\newblock \showarticletitle{Studying up: reorienting the study of algorithmic
  fairness around issues of power}. In \bibinfo{booktitle}{\emph{Proceedings of
  the 2020 Conference on Fairness, Accountability, and Transparency}}.
  \bibinfo{pages}{167--176}.
\newblock


\bibitem[Barocas et~al\mbox{.}(2019)]%
        {barocas-hardt-narayanan}
\bibfield{author}{\bibinfo{person}{Solon Barocas}, \bibinfo{person}{Moritz
  Hardt}, {and} \bibinfo{person}{Arvind Narayanan}.}
  \bibinfo{year}{2019}\natexlab{}.
\newblock \bibinfo{booktitle}{\emph{Fairness and Machine Learning}}.
\newblock \bibinfo{publisher}{fairmlbook.org}.
\newblock
\newblock
\shownote{\url{http://www.fairmlbook.org}}.


\bibitem[Beyer and Holtzblatt(1999)]%
        {beyer1999contextual}
\bibfield{author}{\bibinfo{person}{Hugh Beyer} {and} \bibinfo{person}{Karen
  Holtzblatt}.} \bibinfo{year}{1999}\natexlab{}.
\newblock \showarticletitle{Contextual design}.
\newblock \bibinfo{journal}{\emph{interactions}} \bibinfo{volume}{6},
  \bibinfo{number}{1} (\bibinfo{year}{1999}), \bibinfo{pages}{32--42}.
\newblock


\bibitem[Braun and Clarke(2019)]%
        {braun2019reflecting}
\bibfield{author}{\bibinfo{person}{Virginia Braun} {and}
  \bibinfo{person}{Victoria Clarke}.} \bibinfo{year}{2019}\natexlab{}.
\newblock \showarticletitle{Reflecting on reflexive thematic analysis}.
\newblock \bibinfo{journal}{\emph{Qualitative research in sport, exercise and
  health}} \bibinfo{volume}{11}, \bibinfo{number}{4} (\bibinfo{year}{2019}),
  \bibinfo{pages}{589--597}.
\newblock


\bibitem[Brown et~al\mbox{.}(2019)]%
        {Brown2019}
\bibfield{author}{\bibinfo{person}{Anna Brown}, \bibinfo{person}{Alexandra
  Chouldechova}, \bibinfo{person}{Emily Putnam-Hornstein},
  \bibinfo{person}{Andrew Tobin}, {and} \bibinfo{person}{Rhema Vaithianathan}.}
  \bibinfo{year}{2019}\natexlab{}.
\newblock \showarticletitle{Toward algorithmic accountability in public
  services: A qualitative study of affected community perspectives on
  algorithmic decision-making in child welfare services}. In
  \bibinfo{booktitle}{\emph{Proceedings of the 2019 CHI Conference on Human
  Factors in Computing Systems}}. \bibinfo{pages}{1--12}.
\newblock


\bibitem[Cheng et~al\mbox{.}(2022)]%
        {cheng2022child}
\bibfield{author}{\bibinfo{person}{Hao-Fei Cheng}, \bibinfo{person}{Logan
  Stapleton}, \bibinfo{person}{Anna Kawakami}, \bibinfo{person}{Venkatesh
  Sivaraman}, \bibinfo{person}{Yanghuidi Cheng}, \bibinfo{person}{Diana Qing},
  \bibinfo{person}{Adam Perer}, \bibinfo{person}{Kenneth Holstein},
  \bibinfo{person}{Zhiwei~Steven Wu}, {and} \bibinfo{person}{Haiyi Zhu}.}
  \bibinfo{year}{2022}\natexlab{}.
\newblock \showarticletitle{How Child Welfare Workers Reduce Racial Disparities
  in Algorithmic Decisions}. In \bibinfo{booktitle}{\emph{CHI Conference on
  Human Factors in Computing Systems}}. \bibinfo{pages}{1--22}.
\newblock


\bibitem[Chouldechova et~al\mbox{.}(2018)]%
        {chouldechova2018case}
\bibfield{author}{\bibinfo{person}{Alexandra Chouldechova},
  \bibinfo{person}{Diana Benavides-Prado}, \bibinfo{person}{Oleksandr Fialko},
  {and} \bibinfo{person}{Rhema Vaithianathan}.}
  \bibinfo{year}{2018}\natexlab{}.
\newblock \showarticletitle{A case study of algorithm-assisted decision making
  in child maltreatment hotline screening decisions}. In
  \bibinfo{booktitle}{\emph{Conference on Fairness, Accountability and
  Transparency}}. PMLR, \bibinfo{pages}{134--148}.
\newblock


\bibitem[Coston et~al\mbox{.}(2022)]%
        {coston2022validity}
\bibfield{author}{\bibinfo{person}{Amanda Coston}, \bibinfo{person}{Anna
  Kawakami}, \bibinfo{person}{Haiyi Zhu}, \bibinfo{person}{Ken Holstein}, {and}
  \bibinfo{person}{Hoda Heidari}.} \bibinfo{year}{2022}\natexlab{}.
\newblock \showarticletitle{A Validity Perspective on Evaluating the Justified
  Use of Data-driven Decision-making Algorithms}.
\newblock \bibinfo{journal}{\emph{arXiv preprint arXiv:2206.14983}}
  (\bibinfo{year}{2022}).
\newblock


\bibitem[Crawford(2021)]%
        {crawford2021atlas}
\bibfield{author}{\bibinfo{person}{Kate Crawford}.}
  \bibinfo{year}{2021}\natexlab{}.
\newblock \bibinfo{booktitle}{\emph{The atlas of AI: Power, politics, and the
  planetary costs of artificial intelligence}}.
\newblock \bibinfo{publisher}{Yale University Press}.
\newblock


\bibitem[Delgado et~al\mbox{.}(2021)]%
        {delgado2021stakeholder}
\bibfield{author}{\bibinfo{person}{Fernando Delgado}, \bibinfo{person}{Stephen
  Yang}, \bibinfo{person}{Michael Madaio}, {and} \bibinfo{person}{Qian Yang}.}
  \bibinfo{year}{2021}\natexlab{}.
\newblock \showarticletitle{Stakeholder Participation in AI: Beyond" Add
  Diverse Stakeholders and Stir"}.
\newblock \bibinfo{journal}{\emph{arXiv preprint arXiv:2111.01122}}
  (\bibinfo{year}{2021}).
\newblock


\bibitem[Dow et~al\mbox{.}(2018)]%
        {dow2018between}
\bibfield{author}{\bibinfo{person}{Andy Dow}, \bibinfo{person}{Rob Comber},
  {and} \bibinfo{person}{John Vines}.} \bibinfo{year}{2018}\natexlab{}.
\newblock \showarticletitle{Between grassroots and the hierarchy: Lessons
  learned from the design of a public services directory}. In
  \bibinfo{booktitle}{\emph{Proceedings of the 2018 CHI Conference on Human
  Factors in Computing Systems}}. \bibinfo{pages}{1--13}.
\newblock


\bibitem[Eubanks(2018)]%
        {eubanks2018automating}
\bibfield{author}{\bibinfo{person}{Virginia Eubanks}.}
  \bibinfo{year}{2018}\natexlab{}.
\newblock \bibinfo{booktitle}{\emph{Automating inequality: How high-tech tools
  profile, police, and punish the poor}}.
\newblock \bibinfo{publisher}{St. Martin's Press}.
\newblock


\bibitem[Forsythe(2001)]%
        {forsythe2001studying}
\bibfield{author}{\bibinfo{person}{Diana Forsythe}.}
  \bibinfo{year}{2001}\natexlab{}.
\newblock \bibinfo{booktitle}{\emph{Studying those who study us: An
  anthropologist in the world of artificial intelligence}}.
\newblock \bibinfo{publisher}{Stanford University Press}.
\newblock


\bibitem[Gerchick et~al\mbox{.}(2023)]%
        {gerchick2023devil}
\bibfield{author}{\bibinfo{person}{Marissa Gerchick}, \bibinfo{person}{Tobi
  Jegede}, \bibinfo{person}{Tarak Shah}, \bibinfo{person}{Ana Gutierrez},
  \bibinfo{person}{Sophie Beiers}, \bibinfo{person}{Noam Shemtov},
  \bibinfo{person}{Kath Xu}, \bibinfo{person}{Anjana Samant}, {and}
  \bibinfo{person}{Aaron Horowitz}.} \bibinfo{year}{2023}\natexlab{}.
\newblock \showarticletitle{The Devil is in the Details: Interrogating Values
  Embedded in the Allegheny Family Screening Tool}. In
  \bibinfo{booktitle}{\emph{Proceedings of the 2023 ACM Conference on Fairness,
  Accountability, and Transparency}}. \bibinfo{pages}{1292--1310}.
\newblock


\bibitem[Goldhaber-Fiebert and Prince(2019)]%
        {goldhaber2019impact}
\bibfield{author}{\bibinfo{person}{Jeremy~D Goldhaber-Fiebert} {and}
  \bibinfo{person}{Lea Prince}.} \bibinfo{year}{2019}\natexlab{}.
\newblock \showarticletitle{Impact evaluation of a predictive risk modeling
  tool for Allegheny county’s child welfare office}.
\newblock \bibinfo{journal}{\emph{Pittsburgh: Allegheny County}}
  (\bibinfo{year}{2019}).
\newblock


\bibitem[Holstein and Doroudi(2019)]%
        {holstein2019fairness}
\bibfield{author}{\bibinfo{person}{Kenneth Holstein} {and}
  \bibinfo{person}{Shayan Doroudi}.} \bibinfo{year}{2019}\natexlab{}.
\newblock \showarticletitle{Fairness and equity in learning analytics systems
  (FairLAK)}. In \bibinfo{booktitle}{\emph{Companion proceedings of the ninth
  international learning analytics \& knowledge conference (LAK 2019)}}.
  \bibinfo{pages}{1--2}.
\newblock


\bibitem[Holstein et~al\mbox{.}(2017)]%
        {holstein2017intelligent}
\bibfield{author}{\bibinfo{person}{Kenneth Holstein}, \bibinfo{person}{Bruce~M
  McLaren}, {and} \bibinfo{person}{Vincent Aleven}.}
  \bibinfo{year}{2017}\natexlab{}.
\newblock \showarticletitle{Intelligent tutors as teachers' aides: exploring
  teacher needs for real-time analytics in blended classrooms}. In
  \bibinfo{booktitle}{\emph{Proceedings of the seventh international learning
  analytics \& knowledge conference}}. \bibinfo{pages}{257--266}.
\newblock


\bibitem[Holstein et~al\mbox{.}(2018)]%
        {holstein2018student}
\bibfield{author}{\bibinfo{person}{Kenneth Holstein}, \bibinfo{person}{Bruce~M
  McLaren}, {and} \bibinfo{person}{Vincent Aleven}.}
  \bibinfo{year}{2018}\natexlab{}.
\newblock \showarticletitle{Student learning benefits of a mixed-reality
  teacher awareness tool in AI-enhanced classrooms}. In
  \bibinfo{booktitle}{\emph{International conference on artificial intelligence
  in education}}. Springer, \bibinfo{pages}{154--168}.
\newblock


\bibitem[{Holten M{\o}ller} et~al\mbox{.}(2020)]%
        {HoltenMoller2020}
\bibfield{author}{\bibinfo{person}{Naja {Holten M{\o}ller}},
  \bibinfo{person}{Irina Shklovski}, {and} \bibinfo{person}{Thomas~T.
  Hildebrandt}.} \bibinfo{year}{2020}\natexlab{}.
\newblock \showarticletitle{{Shifting concepts of value: Designing algorithmic
  decision-support systems for public services}}.
\newblock \bibinfo{journal}{\emph{NordiCHI}} (\bibinfo{year}{2020}),
  \bibinfo{pages}{1--12}.
\newblock
\showISBNx{9781450375795}
\urldef\tempurl%
\url{https://doi.org/10.1145/3419249.3420149}
\showDOI{\tempurl}


\bibitem[House(2023)]%
        {house2023fact}
\bibfield{author}{\bibinfo{person}{White House}.}
  \bibinfo{year}{2023}\natexlab{}.
\newblock \bibinfo{title}{Fact sheet: President Biden issues executive order on
  safe, secure, and trustworthy artificial intelligence}.
\newblock
\newblock


\bibitem[Kang and Fox(2022)]%
        {kang2022stories}
\bibfield{author}{\bibinfo{person}{Esther~Y Kang} {and}
  \bibinfo{person}{Sarah~E Fox}.} \bibinfo{year}{2022}\natexlab{}.
\newblock \showarticletitle{Stories from the Frontline: Recuperating Essential
  Worker Accounts of AI Integration}. In \bibinfo{booktitle}{\emph{Designing
  Interactive Systems Conference}}. \bibinfo{pages}{58--70}.
\newblock


\bibitem[Kawakami et~al\mbox{.}(2024)]%
        {kawakami2024situate}
\bibfield{author}{\bibinfo{person}{Anna Kawakami}, \bibinfo{person}{Amanda
  Coston}, \bibinfo{person}{Haiyi Zhu}, \bibinfo{person}{Hoda Heidari}, {and}
  \bibinfo{person}{Kenneth Holstein}.} \bibinfo{year}{2024}\natexlab{}.
\newblock \showarticletitle{The Situate AI Guidebook: Co-Designing a Toolkit to
  Support Multi-Stakeholder, Early-stage Deliberations Around Public Sector AI
  Proposals}. In \bibinfo{booktitle}{\emph{Proceedings of the CHI Conference on
  Human Factors in Computing Systems}}. \bibinfo{pages}{1--22}.
\newblock


\bibitem[Kawakami et~al\mbox{.}(2022a)]%
        {kawakami2022improving}
\bibfield{author}{\bibinfo{person}{Anna Kawakami}, \bibinfo{person}{Venkatesh
  Sivaraman}, \bibinfo{person}{Hao-Fei Cheng}, \bibinfo{person}{Logan
  Stapleton}, \bibinfo{person}{Yanghuidi Cheng}, \bibinfo{person}{Diana Qing},
  \bibinfo{person}{Adam Perer}, \bibinfo{person}{Zhiwei~Steven Wu},
  \bibinfo{person}{Haiyi Zhu}, {and} \bibinfo{person}{Kenneth Holstein}.}
  \bibinfo{year}{2022}\natexlab{a}.
\newblock \showarticletitle{Improving Human-AI Partnerships in Child Welfare:
  Understanding Worker Practices, Challenges, and Desires for Algorithmic
  Decision Support}. In \bibinfo{booktitle}{\emph{CHI Conference on Human
  Factors in Computing Systems}}. \bibinfo{pages}{1--18}.
\newblock


\bibitem[Kawakami et~al\mbox{.}(2022b)]%
        {kawakami2022care}
\bibfield{author}{\bibinfo{person}{Anna Kawakami}, \bibinfo{person}{Venkatesh
  Sivaraman}, \bibinfo{person}{Logan Stapleton}, \bibinfo{person}{Hao-Fei
  Cheng}, \bibinfo{person}{Adam Perer}, \bibinfo{person}{Zhiwei~Steven Wu},
  \bibinfo{person}{Haiyi Zhu}, {and} \bibinfo{person}{Kenneth Holstein}.}
  \bibinfo{year}{2022}\natexlab{b}.
\newblock \showarticletitle{“Why Do I Care What’s Similar?” Probing
  Challenges in AI-Assisted Child Welfare Decision-Making through Worker-AI
  Interface Design Concepts}. In \bibinfo{booktitle}{\emph{Designing
  Interactive Systems Conference}}. \bibinfo{pages}{454--470}.
\newblock


\bibitem[Kitchin and Lauriault(2014)]%
        {kitchin2014towards}
\bibfield{author}{\bibinfo{person}{Rob Kitchin} {and} \bibinfo{person}{Tracey
  Lauriault}.} \bibinfo{year}{2014}\natexlab{}.
\newblock \showarticletitle{Towards critical data studies: Charting and
  unpacking data assemblages and their work}.
\newblock  (\bibinfo{year}{2014}).
\newblock


\bibitem[Kulynych et~al\mbox{.}(2020)]%
        {kulynych2020participatory}
\bibfield{author}{\bibinfo{person}{Bogdan Kulynych}, \bibinfo{person}{David
  Madras}, \bibinfo{person}{Smitha Milli}, \bibinfo{person}{Inioluwa~Deborah
  Raji}, \bibinfo{person}{Angela Zhou}, {and} \bibinfo{person}{Richard Zemel}.}
  \bibinfo{year}{2020}\natexlab{}.
\newblock \showarticletitle{Participatory approaches to machine learning}. In
  \bibinfo{booktitle}{\emph{International Conference on Machine Learning
  Workshop}}, Vol.~\bibinfo{volume}{7}.
\newblock


\bibitem[Kuo et~al\mbox{.}(2023)]%
        {kuo2023understanding}
\bibfield{author}{\bibinfo{person}{Tzu-Sheng Kuo}, \bibinfo{person}{Hong Shen},
  \bibinfo{person}{Jisoo Geum}, \bibinfo{person}{Nev Jones},
  \bibinfo{person}{Jason~I Hong}, \bibinfo{person}{Haiyi Zhu}, {and}
  \bibinfo{person}{Kenneth Holstein}.} \bibinfo{year}{2023}\natexlab{}.
\newblock \showarticletitle{Understanding Frontline Workers’ and Unhoused
  Individuals’ Perspectives on AI Used in Homeless Services}. In
  \bibinfo{booktitle}{\emph{Proceedings of the 2023 CHI Conference on Human
  Factors in Computing Systems}}. \bibinfo{pages}{1--17}.
\newblock


\bibitem[Kuziemski and Misuraca(2020)]%
        {kuziemski2020ai}
\bibfield{author}{\bibinfo{person}{Maciej Kuziemski} {and}
  \bibinfo{person}{Gianluca Misuraca}.} \bibinfo{year}{2020}\natexlab{}.
\newblock \showarticletitle{AI governance in the public sector: Three tales
  from the frontiers of automated decision-making in democratic settings}.
\newblock \bibinfo{journal}{\emph{Telecommunications policy}}
  \bibinfo{volume}{44}, \bibinfo{number}{6} (\bibinfo{year}{2020}),
  \bibinfo{pages}{101976}.
\newblock


\bibitem[Levy et~al\mbox{.}(2021a)]%
        {Levy2021}
\bibfield{author}{\bibinfo{person}{Karen Levy}, \bibinfo{person}{Kyla~E
  Chasalow}, {and} \bibinfo{person}{Sarah Riley}.}
  \bibinfo{year}{2021}\natexlab{a}.
\newblock \showarticletitle{{Algorithms and Decision-Making in the Public
  Sector}}.
\newblock \bibinfo{journal}{\emph{Annual Review of Law and Social Science}}
  \bibinfo{volume}{17} (\bibinfo{year}{2021}), \bibinfo{pages}{1--38}.
\newblock


\bibitem[Levy et~al\mbox{.}(2021b)]%
        {levy2021algorithms}
\bibfield{author}{\bibinfo{person}{Karen Levy}, \bibinfo{person}{Kyla~E
  Chasalow}, {and} \bibinfo{person}{Sarah Riley}.}
  \bibinfo{year}{2021}\natexlab{b}.
\newblock \showarticletitle{Algorithms and decision-making in the public
  sector}.
\newblock \bibinfo{journal}{\emph{Annual Review of Law and Social Science}}
  \bibinfo{volume}{17} (\bibinfo{year}{2021}), \bibinfo{pages}{309--334}.
\newblock


\bibitem[Loi and Spielkamp(2021)]%
        {loi2021towards}
\bibfield{author}{\bibinfo{person}{Michele Loi} {and} \bibinfo{person}{Matthias
  Spielkamp}.} \bibinfo{year}{2021}\natexlab{}.
\newblock \showarticletitle{Towards accountability in the use of artificial
  intelligence for public administrations}. In
  \bibinfo{booktitle}{\emph{Proceedings of the 2021 AAAI/ACM Conference on AI,
  Ethics, and Society}}. \bibinfo{pages}{757--766}.
\newblock


\bibitem[Miceli et~al\mbox{.}(2022)]%
        {miceli2022studying}
\bibfield{author}{\bibinfo{person}{Milagros Miceli}, \bibinfo{person}{Julian
  Posada}, {and} \bibinfo{person}{Tianling Yang}.}
  \bibinfo{year}{2022}\natexlab{}.
\newblock \showarticletitle{Studying up machine learning data: Why talk about
  bias when we mean power?}
\newblock \bibinfo{journal}{\emph{Proceedings of the ACM on Human-Computer
  Interaction}} \bibinfo{volume}{6}, \bibinfo{number}{GROUP}
  (\bibinfo{year}{2022}), \bibinfo{pages}{1--14}.
\newblock


\bibitem[Nader(1972)]%
        {nader1972up}
\bibfield{author}{\bibinfo{person}{Laura Nader}.}
  \bibinfo{year}{1972}\natexlab{}.
\newblock \showarticletitle{Up the anthropologist: Perspectives gained from
  studying up.}
\newblock  (\bibinfo{year}{1972}).
\newblock


\bibitem[Obermeyer et~al\mbox{.}(2019)]%
        {obermeyer2019dissecting}
\bibfield{author}{\bibinfo{person}{Ziad Obermeyer}, \bibinfo{person}{Brian
  Powers}, \bibinfo{person}{Christine Vogeli}, {and} \bibinfo{person}{Sendhil
  Mullainathan}.} \bibinfo{year}{2019}\natexlab{}.
\newblock \showarticletitle{Dissecting racial bias in an algorithm used to
  manage the health of populations}.
\newblock \bibinfo{journal}{\emph{Science}} \bibinfo{volume}{366},
  \bibinfo{number}{6464} (\bibinfo{year}{2019}), \bibinfo{pages}{447--453}.
\newblock


\bibitem[Passi and Barocas(2019)]%
        {passi2019problem}
\bibfield{author}{\bibinfo{person}{Samir Passi} {and} \bibinfo{person}{Solon
  Barocas}.} \bibinfo{year}{2019}\natexlab{}.
\newblock \showarticletitle{Problem formulation and fairness}. In
  \bibinfo{booktitle}{\emph{Proceedings of the conference on fairness,
  accountability, and transparency}}. \bibinfo{pages}{39--48}.
\newblock


\bibitem[Pilemalm(2018)]%
        {pilemalm2018participatory}
\bibfield{author}{\bibinfo{person}{Sofie Pilemalm}.}
  \bibinfo{year}{2018}\natexlab{}.
\newblock \showarticletitle{Participatory design in emerging civic engagement
  initiatives in the new public sector: Applying PD concepts in resource-scarce
  organizations}.
\newblock \bibinfo{journal}{\emph{ACM Transactions on Computer-Human
  Interaction (TOCHI)}} \bibinfo{volume}{25}, \bibinfo{number}{1}
  (\bibinfo{year}{2018}), \bibinfo{pages}{1--26}.
\newblock


\bibitem[Raji et~al\mbox{.}(2022)]%
        {raji2022fallacy}
\bibfield{author}{\bibinfo{person}{Inioluwa~Deborah Raji},
  \bibinfo{person}{I~Elizabeth Kumar}, \bibinfo{person}{Aaron Horowitz}, {and}
  \bibinfo{person}{Andrew Selbst}.} \bibinfo{year}{2022}\natexlab{}.
\newblock \showarticletitle{The fallacy of AI functionality}. In
  \bibinfo{booktitle}{\emph{2022 ACM Conference on Fairness, Accountability,
  and Transparency}}. \bibinfo{pages}{959--972}.
\newblock


\bibitem[Robertson et~al\mbox{.}(2021)]%
        {robertson2021modeling}
\bibfield{author}{\bibinfo{person}{Samantha Robertson}, \bibinfo{person}{Tonya
  Nguyen}, {and} \bibinfo{person}{Niloufar Salehi}.}
  \bibinfo{year}{2021}\natexlab{}.
\newblock \showarticletitle{Modeling assumptions clash with the real world:
  Transparency, equity, and community challenges for student assignment
  algorithms}. In \bibinfo{booktitle}{\emph{Proceedings of the 2021 CHI
  Conference on Human Factors in Computing Systems}}. \bibinfo{pages}{1--14}.
\newblock


\bibitem[Samant et~al\mbox{.}(2021)]%
        {samant2021family}
\bibfield{author}{\bibinfo{person}{Anjana Samant}, \bibinfo{person}{Aaron
  Horowitz}, \bibinfo{person}{Kath Xu}, {and} \bibinfo{person}{Sophie Beiers}.}
  \bibinfo{year}{2021}\natexlab{}.
\newblock \bibinfo{title}{Family surveillance by algorithm: The rapidly
  spreading tools few have heard of. American Civil Liberties Union
  (ACLU)(2021)}.
\newblock
\newblock


\bibitem[Saxena et~al\mbox{.}(2020)]%
        {saxena2020human}
\bibfield{author}{\bibinfo{person}{Devansh Saxena}, \bibinfo{person}{Karla
  Badillo-Urquiola}, \bibinfo{person}{Pamela~J Wisniewski}, {and}
  \bibinfo{person}{Shion Guha}.} \bibinfo{year}{2020}\natexlab{}.
\newblock \showarticletitle{A human-centered review of algorithms used within
  the US child welfare system}. In \bibinfo{booktitle}{\emph{Proceedings of the
  2020 CHI Conference on Human Factors in Computing Systems}}.
  \bibinfo{pages}{1--15}.
\newblock


\bibitem[Saxena et~al\mbox{.}(2021)]%
        {saxena2021framework}
\bibfield{author}{\bibinfo{person}{Devansh Saxena}, \bibinfo{person}{Karla
  Badillo-Urquiola}, \bibinfo{person}{Pamela~J Wisniewski}, {and}
  \bibinfo{person}{Shion Guha}.} \bibinfo{year}{2021}\natexlab{}.
\newblock \showarticletitle{A framework of high-stakes algorithmic
  decision-making for the public sector developed through a case study of
  child-welfare}.
\newblock \bibinfo{journal}{\emph{Proceedings of the ACM on Human-Computer
  Interaction}} \bibinfo{volume}{5}, \bibinfo{number}{CSCW2}
  (\bibinfo{year}{2021}), \bibinfo{pages}{1--41}.
\newblock


\bibitem[Seaver(2014)]%
        {seaver2014studying}
\bibfield{author}{\bibinfo{person}{Nick Seaver}.}
  \bibinfo{year}{2014}\natexlab{}.
\newblock \showarticletitle{Studying up: The ethnography of technologists}.
\newblock \bibinfo{journal}{\emph{Ethnography Matters}}  \bibinfo{volume}{10}
  (\bibinfo{year}{2014}).
\newblock


\bibitem[Shapiro(2017)]%
        {shapiro2017reform}
\bibfield{author}{\bibinfo{person}{Aaron Shapiro}.}
  \bibinfo{year}{2017}\natexlab{}.
\newblock \showarticletitle{Reform predictive policing}.
\newblock \bibinfo{journal}{\emph{Nature}} \bibinfo{volume}{541},
  \bibinfo{number}{7638} (\bibinfo{year}{2017}), \bibinfo{pages}{458--460}.
\newblock


\bibitem[Sloane et~al\mbox{.}(2022)]%
        {sloane2022participation}
\bibfield{author}{\bibinfo{person}{Mona Sloane}, \bibinfo{person}{Emanuel
  Moss}, \bibinfo{person}{Olaitan Awomolo}, {and} \bibinfo{person}{Laura
  Forlano}.} \bibinfo{year}{2022}\natexlab{}.
\newblock \showarticletitle{Participation is not a design fix for machine
  learning}. In \bibinfo{booktitle}{\emph{Proceedings of the 2nd ACM Conference
  on Equity and Access in Algorithms, Mechanisms, and Optimization}}.
  \bibinfo{pages}{1--6}.
\newblock


\bibitem[Smith et~al\mbox{.}(2020)]%
        {Smith2020}
\bibfield{author}{\bibinfo{person}{C.~Estelle Smith}, \bibinfo{person}{Bowen
  Yu}, \bibinfo{person}{Anjali Srivastava}, \bibinfo{person}{Aaron Halfaker},
  \bibinfo{person}{Loren Terveen}, {and} \bibinfo{person}{Haiyi Zhu}.}
  \bibinfo{year}{2020}\natexlab{}.
\newblock \showarticletitle{{Keeping Community in the Loop: Understanding
  Wikipedia Stakeholder Values for Machine Learning-Based Systems}}.
\newblock \bibinfo{journal}{\emph{Conference on Human Factors in Computing
  Systems - Proceedings}} (\bibinfo{year}{2020}), \bibinfo{pages}{1--14}.
\newblock
\showISBNx{9781450367080}
\urldef\tempurl%
\url{https://doi.org/10.1145/3313831.3376783}
\showDOI{\tempurl}


\bibitem[Stapleton et~al\mbox{.}(2022)]%
        {stapleton2022imagining}
\bibfield{author}{\bibinfo{person}{Logan Stapleton}, \bibinfo{person}{Min~Hun
  Lee}, \bibinfo{person}{Diana Qing}, \bibinfo{person}{Marya Wright},
  \bibinfo{person}{Alexandra Chouldechova}, \bibinfo{person}{Ken Holstein},
  \bibinfo{person}{Zhiwei~Steven Wu}, {and} \bibinfo{person}{Haiyi Zhu}.}
  \bibinfo{year}{2022}\natexlab{}.
\newblock \showarticletitle{Imagining new futures beyond predictive systems in
  child welfare: A qualitative study with impacted stakeholders}. In
  \bibinfo{booktitle}{\emph{2022 ACM Conference on Fairness, Accountability,
  and Transparency}}. \bibinfo{pages}{1162--1177}.
\newblock


\bibitem[Tan et~al\mbox{.}(2018)]%
        {tan2018investigating}
\bibfield{author}{\bibinfo{person}{Sarah Tan}, \bibinfo{person}{Julius
  Adebayo}, \bibinfo{person}{Kori Inkpen}, {and} \bibinfo{person}{Ece Kamar}.}
  \bibinfo{year}{2018}\natexlab{}.
\newblock \showarticletitle{Investigating human+ machine complementarity for
  recidivism predictions}.
\newblock \bibinfo{journal}{\emph{arXiv preprint arXiv:1808.09123}}
  (\bibinfo{year}{2018}).
\newblock


\bibitem[Turner et~al\mbox{.}(2003)]%
        {turner2003turning}
\bibfield{author}{\bibinfo{person}{Mark Turner}, \bibinfo{person}{David
  Budgen}, {and} \bibinfo{person}{Pearl Brereton}.}
  \bibinfo{year}{2003}\natexlab{}.
\newblock \showarticletitle{Turning software into a service}.
\newblock \bibinfo{journal}{\emph{Computer}} \bibinfo{volume}{36},
  \bibinfo{number}{10} (\bibinfo{year}{2003}), \bibinfo{pages}{38--44}.
\newblock


\bibitem[van~den Broek et~al\mbox{.}(2020)]%
        {VandenBroek2020}
\bibfield{author}{\bibinfo{person}{Elmira van~den Broek},
  \bibinfo{person}{Anastasia Sergeeva}, {and} \bibinfo{person}{Marleen
  Huysman}.} \bibinfo{year}{2020}\natexlab{}.
\newblock \showarticletitle{{Hiring algorithms: An ethnography of fairness in
  practice}}.
\newblock \bibinfo{journal}{\emph{40th International Conference on Information
  Systems, ICIS 2019}} (\bibinfo{year}{2020}).
\newblock
\showISBNx{9780996683197}


\bibitem[Veale et~al\mbox{.}(2018)]%
        {veale2018fairness}
\bibfield{author}{\bibinfo{person}{Michael Veale}, \bibinfo{person}{Max
  Van~Kleek}, {and} \bibinfo{person}{Reuben Binns}.}
  \bibinfo{year}{2018}\natexlab{}.
\newblock \showarticletitle{Fairness and accountability design needs for
  algorithmic support in high-stakes public sector decision-making}. In
  \bibinfo{booktitle}{\emph{Proceedings of the 2018 chi conference on human
  factors in computing systems}}. \bibinfo{pages}{1--14}.
\newblock


\bibitem[Wang et~al\mbox{.}(2022)]%
        {wang2022against}
\bibfield{author}{\bibinfo{person}{Angelina Wang}, \bibinfo{person}{Sayash
  Kapoor}, \bibinfo{person}{Solon Barocas}, {and} \bibinfo{person}{Arvind
  Narayanan}.} \bibinfo{year}{2022}\natexlab{}.
\newblock \showarticletitle{Against predictive optimization: On the legitimacy
  of decision-making algorithms that optimize predictive accuracy}.
\newblock \bibinfo{journal}{\emph{Available at SSRN}} (\bibinfo{year}{2022}).
\newblock


\bibitem[Wolf et~al\mbox{.}(2018)]%
        {wolf2018changing}
\bibfield{author}{\bibinfo{person}{Christine~T Wolf}, \bibinfo{person}{Haiyi
  Zhu}, \bibinfo{person}{Julia Bullard}, \bibinfo{person}{Min~Kyung Lee}, {and}
  \bibinfo{person}{Jed~R Brubaker}.} \bibinfo{year}{2018}\natexlab{}.
\newblock \showarticletitle{The changing contours of" participation" in
  data-driven, algorithmic ecosystems: Challenges, tactics, and an agenda}. In
  \bibinfo{booktitle}{\emph{Companion of the 2018 ACM Conference on Computer
  Supported Cooperative Work and Social Computing}}. \bibinfo{pages}{377--384}.
\newblock


\bibitem[Wong et~al\mbox{.}(2023)]%
        {wong2023seeing}
\bibfield{author}{\bibinfo{person}{Richmond~Y Wong}, \bibinfo{person}{Michael~A
  Madaio}, {and} \bibinfo{person}{Nick Merrill}.}
  \bibinfo{year}{2023}\natexlab{}.
\newblock \showarticletitle{Seeing like a toolkit: How toolkits envision the
  work of AI ethics}.
\newblock \bibinfo{journal}{\emph{Proceedings of the ACM on Human-Computer
  Interaction}} \bibinfo{volume}{7}, \bibinfo{number}{CSCW1}
  (\bibinfo{year}{2023}), \bibinfo{pages}{1--27}.
\newblock


\bibitem[Yang et~al\mbox{.}(2019)]%
        {Yang2019}
\bibfield{author}{\bibinfo{person}{Qian Yang}, \bibinfo{person}{Aaron
  Steinfeld}, {and} \bibinfo{person}{John Zimmerman}.}
  \bibinfo{year}{2019}\natexlab{}.
\newblock \showarticletitle{{Unremarkable AI: Fitting intelligent decision
  support into critical, clinical decision-making processes}}.
\newblock \bibinfo{journal}{\emph{Conference on Human Factors in Computing
  Systems - Proceedings}} (\bibinfo{year}{2019}).
\newblock
\showISBNx{9781450359702}
\urldef\tempurl%
\url{https://doi.org/10.1145/3290605.3300468}
\showDOI{\tempurl}
\showeprint[arxiv]{1904.09612}


\bibitem[Yu et~al\mbox{.}(2018)]%
        {yu2018artificial}
\bibfield{author}{\bibinfo{person}{Kun-Hsing Yu}, \bibinfo{person}{Andrew~L
  Beam}, {and} \bibinfo{person}{Isaac~S Kohane}.}
  \bibinfo{year}{2018}\natexlab{}.
\newblock \showarticletitle{Artificial intelligence in healthcare}.
\newblock \bibinfo{journal}{\emph{Nature biomedical engineering}}
  \bibinfo{volume}{2}, \bibinfo{number}{10} (\bibinfo{year}{2018}),
  \bibinfo{pages}{719--731}.
\newblock


\bibitem[Zhang et~al\mbox{.}(2023)]%
        {zhang2023deliberating}
\bibfield{author}{\bibinfo{person}{Angie Zhang}, \bibinfo{person}{Olympia
  Walker}, \bibinfo{person}{Kaci Nguyen}, \bibinfo{person}{Jiajun Dai},
  \bibinfo{person}{Anqing Chen}, {and} \bibinfo{person}{Min~Kyung Lee}.}
  \bibinfo{year}{2023}\natexlab{}.
\newblock \showarticletitle{Deliberating with AI: Improving Decision-Making for
  the Future through Participatory AI Design and Stakeholder Deliberation}.
\newblock \bibinfo{journal}{\emph{Proceedings of the ACM on Human-Computer
  Interaction}} \bibinfo{volume}{7}, \bibinfo{number}{CSCW1}
  (\bibinfo{year}{2023}), \bibinfo{pages}{1--32}.
\newblock


\bibitem[Zhu et~al\mbox{.}(2018)]%
        {zhu2018value}
\bibfield{author}{\bibinfo{person}{Haiyi Zhu}, \bibinfo{person}{Bowen Yu},
  \bibinfo{person}{Aaron Halfaker}, {and} \bibinfo{person}{Loren Terveen}.}
  \bibinfo{year}{2018}\natexlab{}.
\newblock \showarticletitle{Value-sensitive algorithm design: Method, case
  study, and lessons}.
\newblock \bibinfo{journal}{\emph{Proceedings of the ACM on Human-Computer
  Interaction}} \bibinfo{volume}{2}, \bibinfo{number}{CSCW}
  (\bibinfo{year}{2018}), \bibinfo{pages}{1--23}.
\newblock


\bibitem[Zytek et~al\mbox{.}(2021)]%
        {zytek2021sibyl}
\bibfield{author}{\bibinfo{person}{Alexandra Zytek}, \bibinfo{person}{Dongyu
  Liu}, \bibinfo{person}{Rhema Vaithianathan}, {and} \bibinfo{person}{Kalyan
  Veeramachaneni}.} \bibinfo{year}{2021}\natexlab{}.
\newblock \showarticletitle{Sibyl: Understanding and addressing the usability
  challenges of machine learning in high-stakes decision making}.
\newblock \bibinfo{journal}{\emph{IEEE Transactions on Visualization and
  Computer Graphics}} (\bibinfo{year}{2021}).
\newblock


\end{thebibliography}

% \appendix


\end{document}